\documentclass{aa}

\usepackage{graphicx}
\usepackage{natbib}
\bibpunct{(}{)}{;}{a}{}{,}

\newcommand{\GILDAS}{\texttt{GILDAS}}

\newcommand{\IRAM}{\textrm{IRAM}}
\newcommand{\IRAMthm}{\textrm{IRAM-30m}}
\newcommand{\PdBI}{\textrm{PdBI}}
\newcommand{\CSO}{\textrm{CSO}}

\newcommand{\ie} {{\em i.e.}}
\newcommand{\eg} {{\em e.g.}}

\newcommand{\HH}   {\mbox{H$_2$}}       
\newcommand{\HHexc}{\mbox{$[$H$_2^\star]$}} 
\newcommand{\thCO} {\mbox{$^{13}$CO}}   
\newcommand{\twCO} {\mbox{$^{12}$CO}}   
\newcommand{\CeiO} {\mbox{C$^{18}$O}}   
\newcommand{\CCH}{\mbox{CCH}}
\newcommand{\CCCCH}{\mbox{C$_4$H}}
\newcommand{\cCCCHH}{\mbox{c-C$_3$H$_2$}}
\newcommand{\Cp}   {\mbox{C$^{+}$}}   
\newcommand{\Jone}{\mbox{J=1--0}}
\newcommand{\Jtwo}{\mbox{J=2--1}}

\newcommand{\emm}[1]{\ensuremath{#1}}   
\newcommand{\emr}[1]{\emm{\mathrm{#1}}} 
\newcommand{\unit}[1]{\emm{\, \emr{#1}}}
\newcommand{\K}   {\unit{K}}
\newcommand{\mm}  {\unit{mm}}
\newcommand{\m }  {\unit{m}}
\newcommand{\mim} {\unit{\mu m}}
\newcommand{\pccm}{\unit{cm^{-3}}}
\newcommand{\pscm}{\unit{cm^{-2}}}

\newcommand{\Kkms}{\unit{K\,km\,s^{-1}}}
\newcommand{\MHz} {\unit{MHz}}

\newcommand{\pc}    {\unit{pc}}

\newcommand{\kms}   {\unit{km\,s^{-1}}}

\renewcommand{\deg}{\emm{^\circ}}

\newcommand{\Av}{\emm{A_V}}
\newcommand{\Rv}{\emm{R_V}}

\newcommand{\TabObs}{%
  \begin{table*}
    \caption{Observation parameters.}
    \begin{center}
      \begin{tabular}{lrrc}
        \hline
        & \multicolumn{2}{c}{Phase center} & Number of fields \\
        Mosaic 1 &  $\alpha_{2000} = 05^h40^m54.27^s $ & $ \delta_{2000} =
        -02^\circ 28' 00''$ & 7 \\
        Mosaic 2 &  $\alpha_{2000} = 05^h40^m53.00^s $ & $ \delta_{2000} =
        -02^\circ 28' 00'' $ & 4 \\
        \hline
      \end{tabular}
      \medskip{}
      \begin{tabular}{lrcccr}
        \hline
        Molecule \& Line & Frequency  &   Beam & PA     & Noise$^{a}$ & \multicolumn{1}{c}{Obs. date} \\
        & GHz        & arcsec & $\deg$ & \Kkms{}     &     \\
        \hline
        Mosaic 1 \\
        \cCCCHH{} $2_{1,2}-1_{0,1}$     &  85.339 & $6.13 \times 4.75$ & 36 & 3.1$\times 10^{-2}$ & Mar. 2002 \& Apr. 2002 \\
        \CCCCH{}-1 N=9-8,J=19/2-17/2    &  85.634 & $6.11 \times 4.74$ & 36 & 2.6$\times 10^{-2}$ & Mar. 2002 \& Apr. 2002 \\
        \CCCCH{}-2 N=9-8,J=17/2-15/2    &  85.672 & $6.11 \times 4.74$ & 36 & 3.4$\times 10^{-2}$ & Mar. 2002 \& Apr. 2002 \\
        \CCH{}-1 N=1-0, J=3/2-1/2 F=2-1 &  87.316 & $7.24 \times 4.99$ & 54 & 3.4$\times 10^{-2}$ & Dec. 2002 \& Mar. 2003 \\
        \CCH{}-2 N=1-0, J=3/2-1/2 F=1-0 &  87.328 & $7.24 \times 4.99$ & 54 & 2.5$\times 10^{-2}$ & Dec. 2002 \& Mar. 2003 \\
        \CCH{}-3 N=1-0, J=1/2-1/2 F=1-1 &  87.402 & $7.24 \times 4.99$ & 54 & 3.4$\times 10^{-2}$ & Dec. 2002 \& Mar. 2003 \\
        \CCH{}-4 N=1-0, J=1/2-1/2 F=0-1 &  87.407 & $7.24 \times 4.99$ & 54 & 2.3$\times 10^{-2}$ & Dec. 2002 \& Mar. 2003 \\
        \CeiO{} \Jtwo{}                 & 219.560 & $6.54 \times 4.31$ & 65 & 9.8$\times 10^{-2}$ &              Mar. 2003 \\
        \hline
        Mosaic 2 \\
        \twCO{} \Jone{} & 115.271 & $5.95 \times 5.00$ & 65 & 1.2$\times 10^{-1}$ & Nov. 1999 \\
        \twCO{} \Jtwo{} & 230.538 & $2.97 \times 2.47$ & 66 & 1.7$\times 10^{-1}$ & Nov. 1999 \\
        \hline
      \end{tabular}
    \end{center}
    $^{a}$ The noise values quoted here are the noises at the mosaic center
    (Please remember that mosaic noise is inhomogeneous due to primary beam
    correction: It steeply increases at the mosaic edges). Those noise values
    have been computed on 1\kms{} velocity bin.
    \label{tab:pdb}
  \end{table*}}

\newcommand{\TabFlux}{%
  \begin{table}
    \centering
    \caption{Calibrator fluxes in Jy.}
    \begin{tabular}{lrrrrrr}
      \hline
      & \multicolumn{2}{c}{B0420$-$014} & \multicolumn{2}{c}{B0607$-$157} & \multicolumn{2}{c}{B0528$+$134}\\
                 & 3\mm & 1\mm & 3\mm & 1\mm & 3\mm & 1\mm \\
      \hline
      27.11.1999 &      &      &      &      &  3.5 &  1.4 \\
      30.03.2002 &  4.8 &      &  2.3 &      &      &      \\
      16.04.2002 &  4.8 &      &  2.5 &      &      &      \\
      22.04.2002 &  4.8 &      &  2.4 &      &      &      \\
      23.12.2002 & 12.5 &      &  2.6 &      &      &      \\
      18.03.2003 & 12.0 &  7.8 &  2.1 & 0.87 &      &      \\
      26.03.2003 & 12.8 &      &  2.1 &      &      &      \\
      \hline
    \end{tabular}
    \label{tab:fluxes}
  \end{table}}

\newcommand{\TabAbundances}{%
  \begin{table}
    \centering %
    \caption{Molecular column densities and abundances at 3 different
      positions of the PDR named ``Cloud'', ``IR peak'' and ``IR edge''.
      Equatorial offsets refer to the Mosaic~2 map center given in Table
      \ref{tab:pdb}. $(\delta x, \delta y)$ offsets refer to the coordinate
      system defined in Fig.~\ref{fig:maps2}. \HH{} column densities have
      been derived from the 1.2\mm{} dust continuum emission using a dust
      temperature range of 20 to 40\K{} for the ``Cloud'' position and 40
      to 80\K{} for the IR positions. Others column densities used LVG
      models with a representative set of densities and kinetic
      temperature. 1--$\sigma$ uncertainties thus reflect the systematics 
      due to the approximate knowledge of density and kinetic temperature. 
      Abundances are computed with respect to the number of protons, 
      \ie{} $[X] = 0.5\,N(X)/N(\HH)$.}
    \begin{tabular}{lrrrr}
      \hline
              & $\delta$RA & $\delta$Dec & $\delta x$ & $\delta y$ \\
      \hline
      Cloud   &  $+6''$ & $-4''$ &  $+24.9''$ & $-5.3''$ \\
      IR peak &  $-6''$ & $-4''$ &  $+12.2''$ & $-2.4''$ \\
      IR edge & $-12''$ & $-4''$ &   $+7.4''$ & $-1.0''$ \\
      \hline
    \end{tabular}
    \medskip{}
    \begin{tabular}{rlr@{$\pm$}lr@{$\pm$}lr@{$\pm$}l}
      \hline
      Quantity     & Unit           & \multicolumn{2}{c}{Cloud} & \multicolumn{2}{c}{IR peak} & \multicolumn{2}{c}{IR edge}      \\
      \hline
      $S_{1.2\mm}$ & mJy/Beam       &    38 & 2     &   35 & 2      &    12 & 2     \\
      $N(\HH)$     & 10$^{21}$\pscm &   27  & 9.5   & 10.5 & 4      &   3.6 & 1.7   \\ 
      $N(\CeiO)$   & 10$^{15}$\pscm &   5.8 & 0.5   &    4 & 0.5    &     1 & 0.3   \\
      $N(\CCH)$    & 10$^{13}$\pscm &   5.5 & 1     &   30 & 5      &    11 & 3     \\
      $N(\cCCCHH)$ & 10$^{12}$\pscm &   2.3 & 0.7   &   24 & 10     &   9.5 & 5     \\
      $N(\CCCCH)$  & 10$^{12}$\pscm &    20 & 10    &   40 & 10     &    37 & 10    \\
      \hline
    \end{tabular}
    \medskip{}
    \begin{tabular}{rlccc}
      \hline
      Quantity     & Unit       & Cloud & IR peak & IR edge \\
      \hline
      $[\CeiO]$    & 10$^{-7}$  &  1.07 &     1.9 &     1.4 \\
      $[\CCH]$     & 10$^{-8}$  &  0.10 &     1.4 &     1.5 \\
      $[\cCCCHH]$  & 10$^{-10}$ &  0.43 &    11.4 &    13.2 \\
      $[\CCCCH]$   & 10$^{-9}$  &  0.37 &     1.9 &     5.2 \\
      \hline
    \end{tabular}
    \label{tab:abondances}
  \end{table}}

\newcommand{\FigMosaic}{%
\begin{figure}
  \centering \includegraphics[height=\hsize{},angle=270]{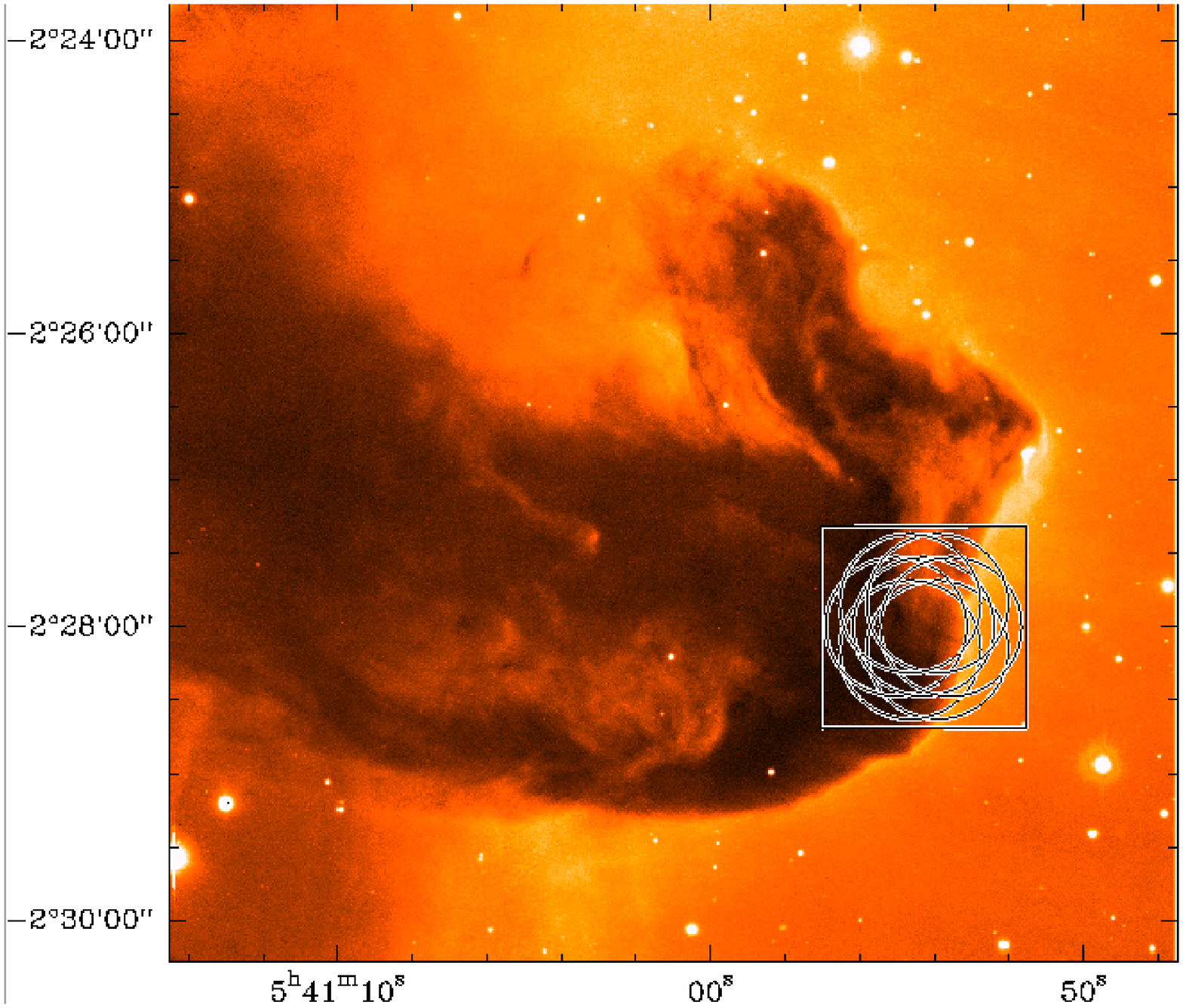}
  \caption{The field of view covered when mapping small hydrocarbons at
    3.4\mm{} with the Plateau de Bure Interferometer (\PdBI{}) is shown as
    a square over this ESO--VLT composite image (B, V and R bands) of the
    Horsehead nebula. Each circle features the 3.4\mm{} primary beam of the
    \PdBI{} at one of the 7 observed positions. Those positions are largely
    oversampled at the hydrocarbon wavelength (3.4\mm{}) to ensure
    simultaneous Nyquist-sampling at 1.4\mm{} used to observe \CeiO{}. A
    linear combination of the 7 pointed observation is done to obtain the
    final dirty image.}
  \label{fig:mosaic}
\end{figure}}

\newcommand{\FigMaps}{%
  \begin{figure*}
    \centering \includegraphics[width=\hsize{}]{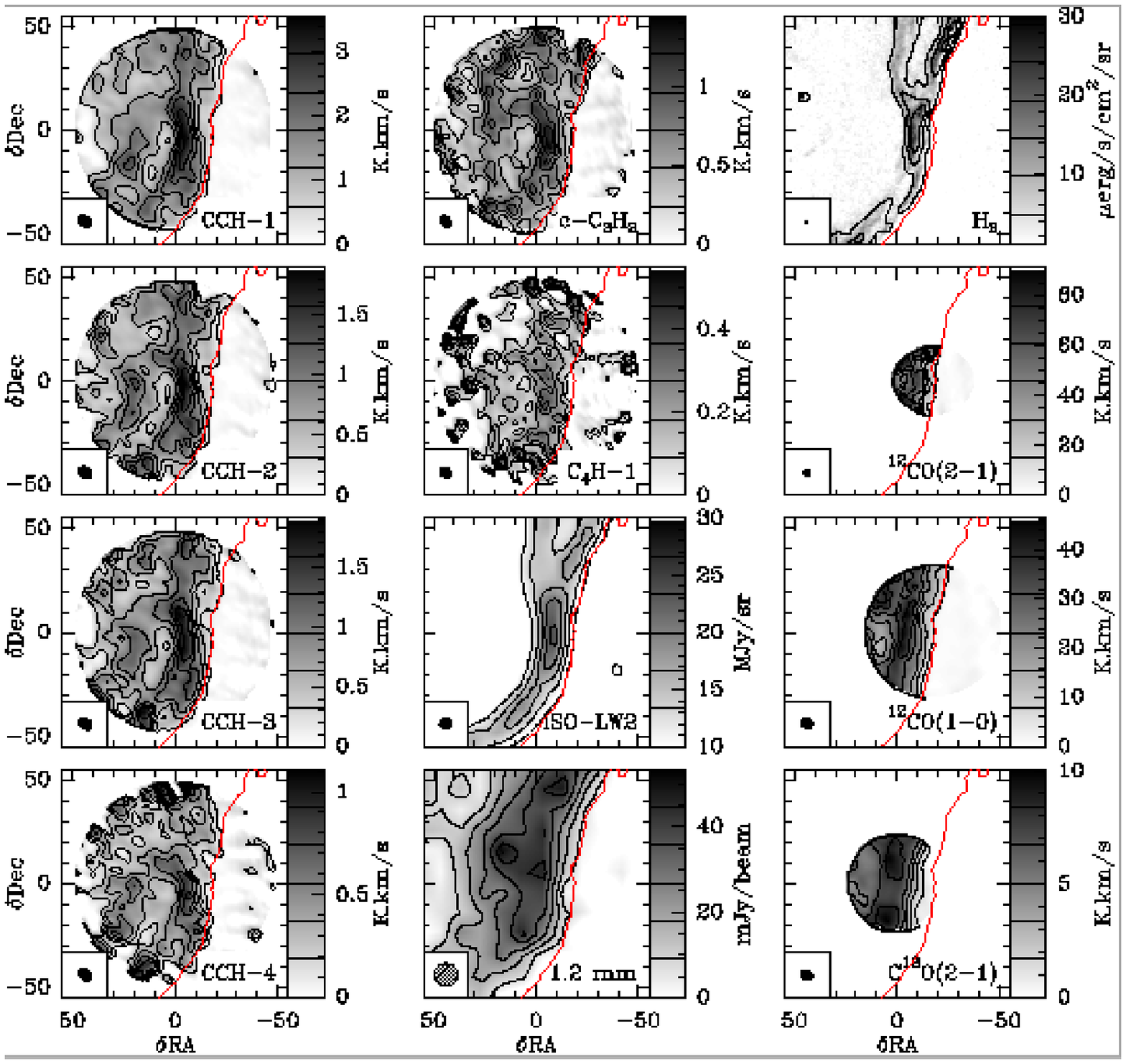}
    \caption{Integrated emission maps obtained with the Plateau de Bure 
      Interferometer. Maps of \emph{i)} the \HH{} v=1-0 S(1)
      emission~\citep{ha04a,ha04b}, \emph{ii)} the mid--IR
      emission~\citep[labeled ISO-LW2]{ab03} and \emph{iii)} the 1.2\mm{}
      dust continuum~\cite[labeled 1.2mm]{tf04} are also shown for
      comparison. The center of all maps has been set to the mosaic 1 phase
      center: RA(2000) = 05h40m54.27s, Dec(2000) = -02\deg{}28'00$''$.  The
      map size is $110'' \times 110''$, with ticks drawn every $10''$.
      Either the synthesized beam or the single dish beam is plotted in the
      bottom, left corner.  The emission of all the lines observed at
      \PdBI{} is integrated between 10.1 and 11.1~\kms{}.  Values of
      contour level are shown on each image wedge (contours of the \HH{}
      image have been computed on an image smoothed to $5''$ resolution).
      The sharp edge of the \HH{} emission (upper right panel) defines a
      boundary, which is used as a numerical support (in the language of
      signal processing) for deconvolution of the other images. This
      deconvolution support is overplotted in red on every panel.}
    \label{fig:maps1}
  \end{figure*}}

\newcommand{\FigMapsRot}{%
  \begin{figure*}
    \centering %
    \includegraphics[width=\hsize{}]{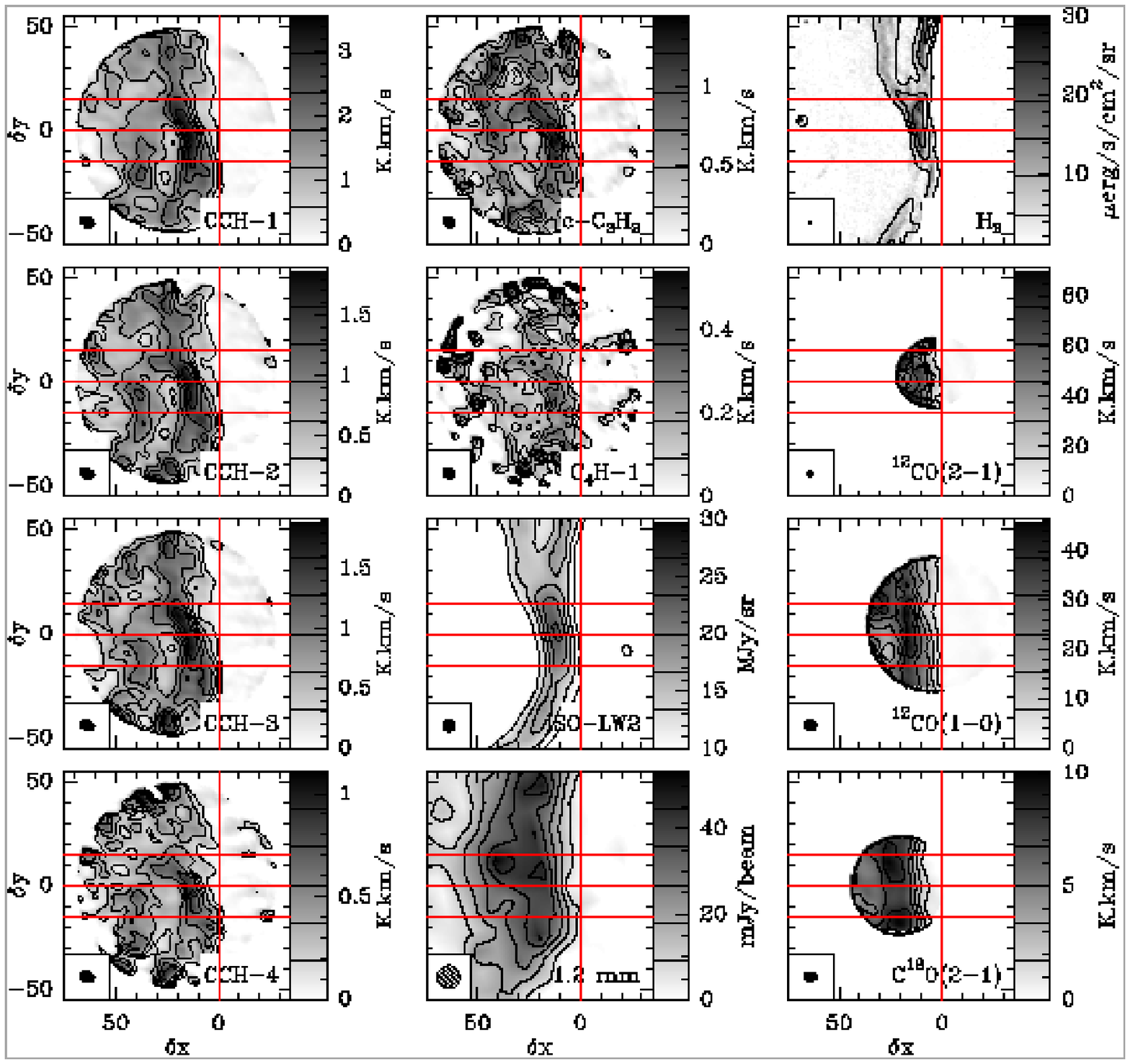}
    \caption{Same as Fig.~\ref{fig:maps1} except that maps have been rotated
      of 14\deg{} counter--clockwise around the image center to bring the
      exciting star direction in the horizontal direction as this eases the
      comparison of the PDR tracer stratifications. Maps have also been
      horizontally shifted by $20''$ to set the horizontal zero at the PDR
      edge delineated as the vertical red line. Horizontal red lines
      delimit the two lanes that have been vertically averaged to produce
      the two series of cuts shown in Fig.~\ref{fig:cuts}.}
    \label{fig:maps2}
  \end{figure*}}

\newcommand{\FigCorrel}{%
\begin{figure}
  \centering %
  \includegraphics[width=0.75\hsize{}]{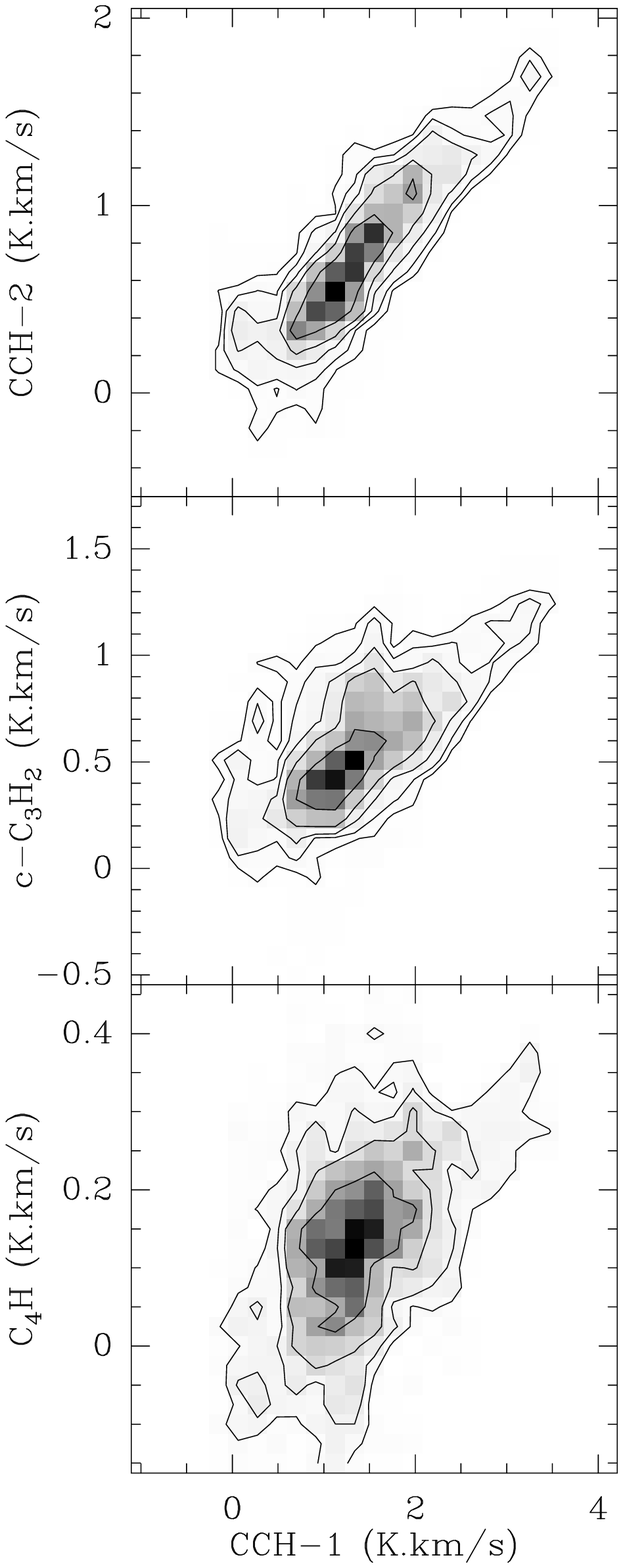}
  \caption{Joint histogram of the integrated emission of \emph{i)} the 
    second brightest \CCH{} line (top), \emph{ii)} \cCCCHH{} (middle) and
    \emph{iii)} one \CCCCH{} line (bottom) vs. the main \CCH{} line.  The
    value at a given position of this joint histogram is the percentage of
    pixels of the input images whose intensities lies in the respective
    vertical and horizontal bins. Only image pixels lying inside the
    deconvolution support (shown in Fig.~\ref{fig:maps1}) have been used in
    the histogram computation.  Contour levels are set to 0.125, 0.25, 0.5,
    1, 2, 4 and 8\% of points per pixel.}
  \label{fig:correl}
\end{figure}}

\newcommand{\FigCuts}{%
  \begin{figure*}
    \centering %
    \includegraphics[width=0.44\hsize{}]{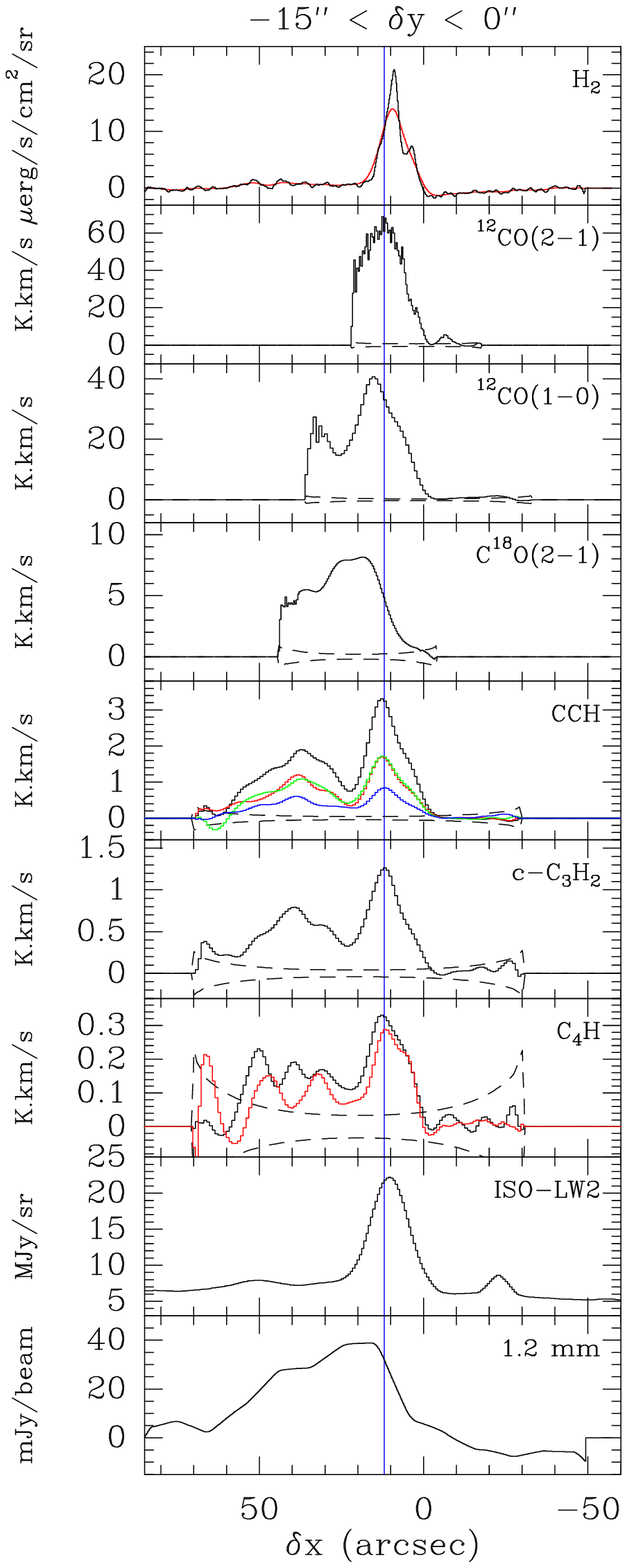}
    \hspace{0.075\hsize}
    \includegraphics[width=0.44\hsize{}]{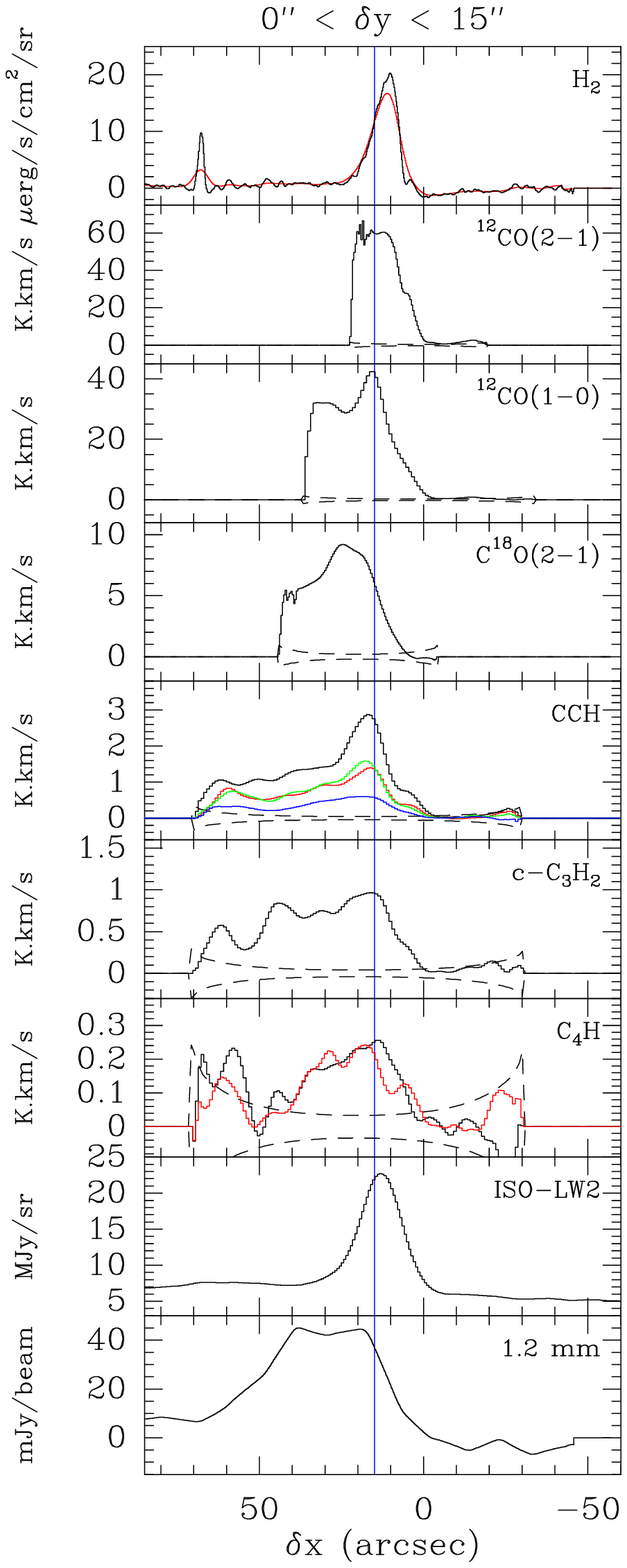}
    \caption{Emission profiles along the exciting star direction (PA =
      -104\deg{} in equatorial coordinate system). To improve
      signal--to--noise ratio, those emission profiles have been integrated
      along the perpendicular direction between $-15'' < \delta y < 0''$
      (left column) and $0'' < \delta y < +15''$ (right column).  The
      comparison of those two series of mean cuts gives an idea of the
      influence of the local conditions (either excitation effects or
      chemical differentiation). We show from top to bottom, \HH{} v=1-0
      S(1) (full resolution in black, smoothed at a $5''$--resolution in
      red), \twCO{} \Jtwo{}, \twCO{} \Jone{}, \CeiO{} \Jtwo{}, \CCH{} lines
      (1 black, 2 green, 3 red and 4 blue, respectively following the
      ratios 1:0.5:0.5:0.25), \cCCCHH{}, \CCCCH{} lines (1 black, 2 red),
      ISO-LW2 and the 1.2\mm{} dust continuum. For the \PdBI{} data, the
      3--$\sigma$ noise level is indicated by the dashed lines. It rises at
      the cut edges due to the primary beam correction.  Note that the
      fields of view of the \twCO{} and \CeiO{} data are smaller than the
      field of view of the hydrocarbon data because of the smaller mosaic
      size and/or the higher frequency. Finally, the blue vertical lines
      give guideline to localize the hydrocarbon peaks. They have been
      drawn at $\delta x =+12''$ (left column) and $\delta x =+15''$ (right
      column).  Note that the bumps at $\delta x = -20''$ in the ISO-LW2
      cut (left column) and at $\delta x =70''$ in the \HH{} cut (right
      column) are due to field stars.}
    \label{fig:cuts}
  \end{figure*}}

\newcommand{\FigCOAnalysis}{%
  \begin{figure}
    \centering %
    \includegraphics[width=\hsize{}]{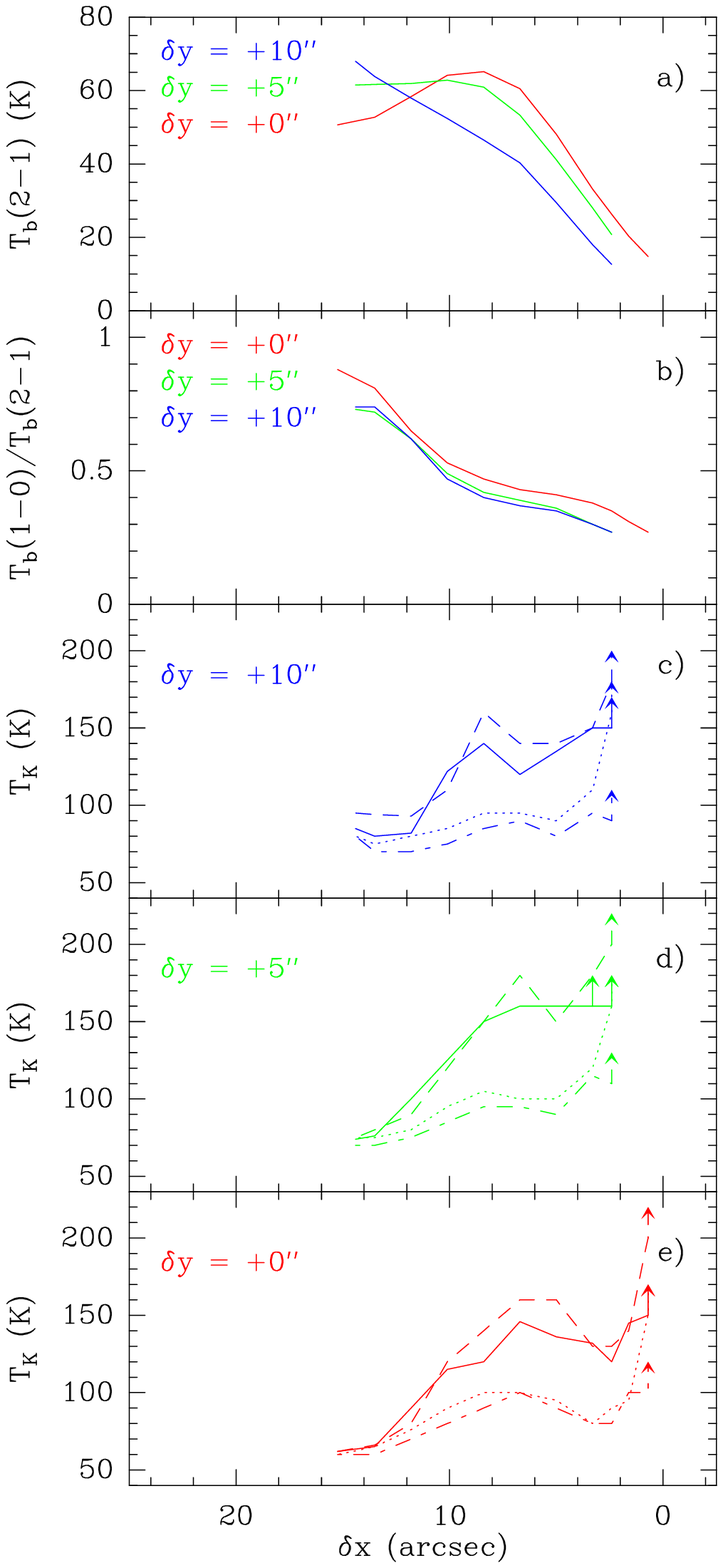}
    \caption{Spatial variation along the direction of the exciting star of 
      a) the \twCO{} \Jtwo{} brightness temperature (convolved at the same
      angular resolution as the \Jone{} transition), b) 1-0/2-1 ratio and
      c,d,e) the kinetic temperature. $(\delta x, \delta y)$ offsets refer
      to the coordinate system defined in Fig.~\ref{fig:maps2}. The cuts at
      $\delta y = +10''$ are drawn in blue, those at $\delta y = +5''$ in
      green and those $\delta y = +0''$ in red. The data cuts have been
      taken from the 10.6\kms{} velocity channel corresponding to the
      \twCO{} line peak. The kinetic temperature is derived from an LVG
      model assuming \emph{i)} unity beam filling factor and \emph{ii)}
      uniform total hydrogen density of $n_\emr{H} = 1.6\times10^4\pccm$
      (dashed lines), $2\times10^4\pccm$ (full lines), $4\times10^4\pccm$
      (dotted lines) and $10^5\pccm$ (dotted--dashed lines). Arrows
      indicate lower limits.}
    \label{fig:co:analysis}
  \end{figure}}

\newcommand{\FigCOSpectra}{%
  \begin{figure*}
    \centering %
    \includegraphics[height=\hsize{},angle=270]{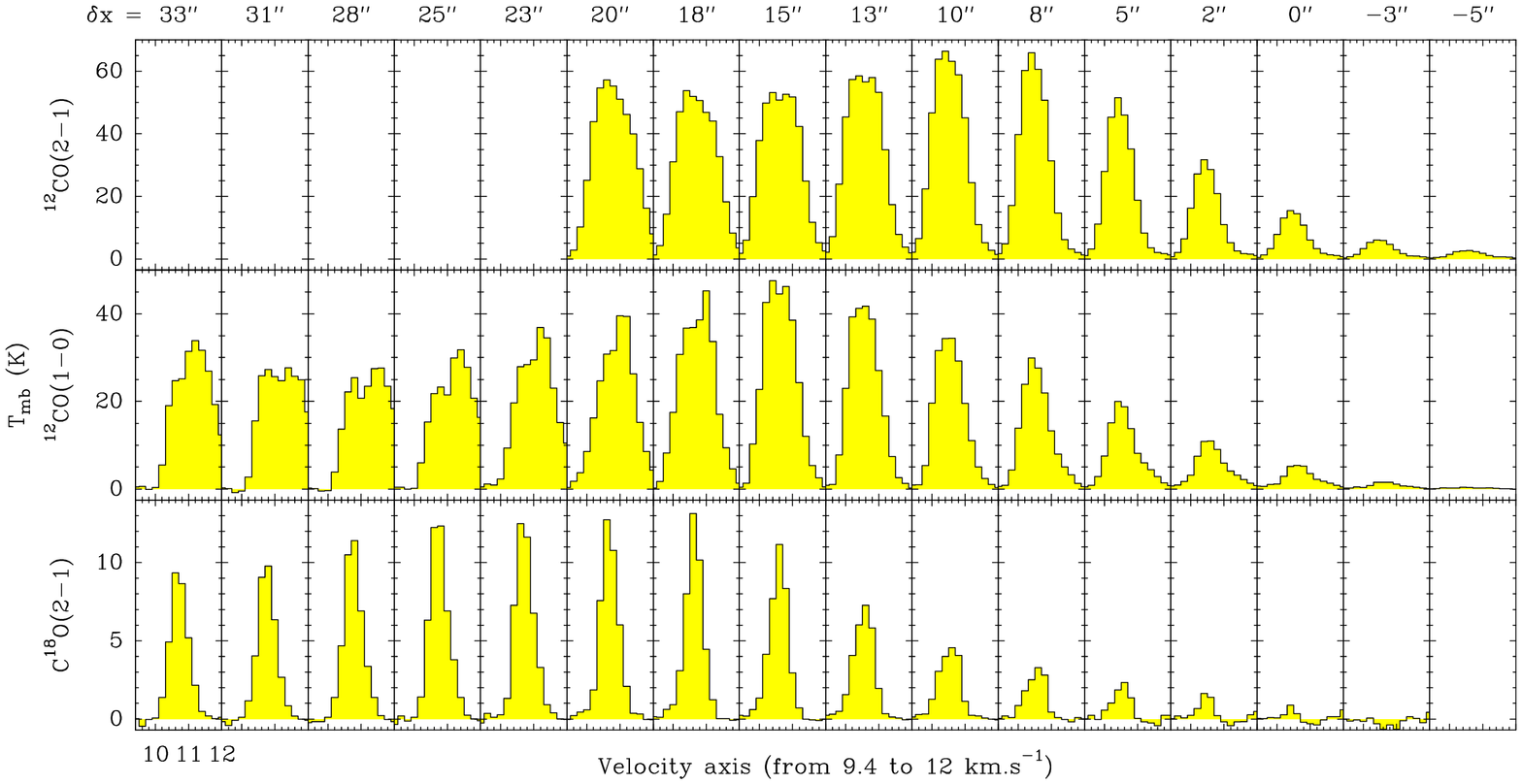}
    \caption{CO spectra (convolved at the same angular resolution)
      along the direction of the exciting star at $\delta y = -2.5''$. In
      those cuts, the label $\delta x = 13''$ indicates the IR peak
      position (cf.~table~\ref{tab:abondances}). Note that \twCO{} \Jtwo{}
      peak intensity decreases at positions $\delta x = 13''$ and $15''$
      while \twCO{} \Jone{} increases even reaching its maximum at $\delta
      x = 15''$.  In addition, both \twCO{} lines show a small but clear
      dip (\ie{} the center channel intensity is lower than its first
      neighbours) at $\delta x = 15''$. Finally, while \CeiO{} spectra are
      very close to Gaussians, \twCO{} spectra show asymmetric profiles.
      Spectra cuts at other close $\delta y$ values show the same trends.}
    \label{fig:co:spectra} 
  \end{figure*}}

\newcommand{\FigModel}{%
  \begin{figure*}
    \centering %
    \includegraphics[width=0.9\hsize{}]{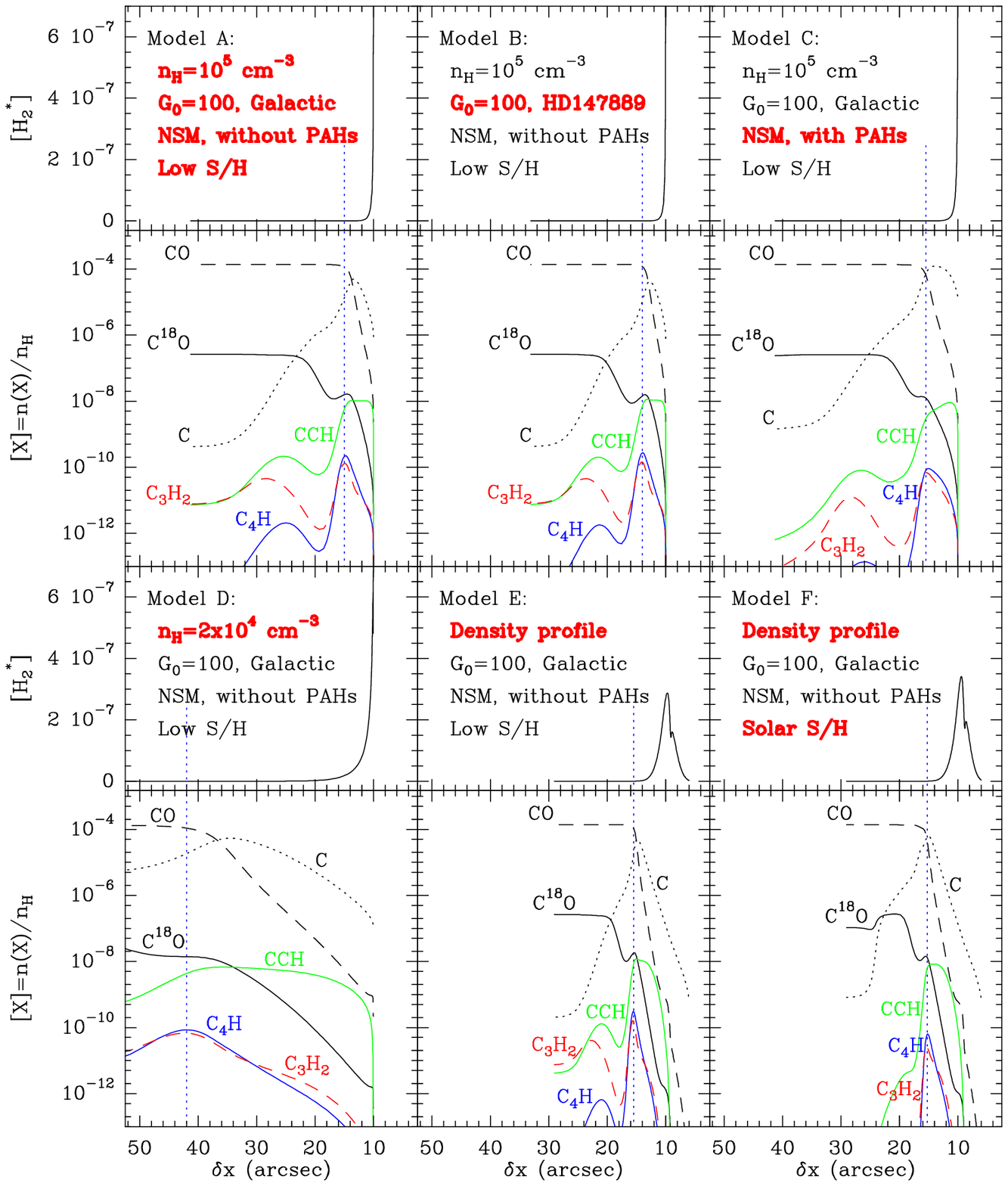}
    \caption{Predictions of the spatial variation of the abundance
      relative to \HH{}, using a unidimensional PDR code. For each model,
      the abundance of the population of the the upper level of the
      2.12\mim{} \HH{} line (\ie{} v=1, J=3), written \HHexc{}, is shown in
      linear scale (top). The C, CO and hydrocarbon abundances are shown in
      logarithmic scale (bottom). The modeled cloud is illuminated from the
      right--hand side.  The $\delta x$--axis origin has been set so that
      \HHexc{} peaks at the position of the observed \HH{} peak (\ie{}
      $\delta x = 10''$).  The vertical dotted blue line indicates the peak
      of the \cCCCHH{} and \CCCCH{} abundances. Each model is described in
      3 lines: \emph{i)} the density structure, \emph{ii)} the UV--field
      properties and \emph{iii)} the chemical network used. 6 different
      models are compared here. Our reference model is labeled A.  The
      total hydrogen density is kept uniform at a value of $n_\emr{H} =
      10^5\pccm$.  The far--UV intensity of the radiation field is
      G$_0=100$ (in Draine units) and the extinction curve is the mean
      Galactic one.  The chemical network rate file is the New Standard
      Model one with minor modifications described in the text.  Charge
      exchange reactions between \Cp{} and PAHs are not taken into account.
      The gaseous sulfur abundance is low compared to solar (\ie{} S/H =
      $5.8\times10^{-8}$). The parameters varied in other models are
      emphasized in red.  Model B uses a different extinction curve.  Model
      C adds reactions of charge exchange between \Cp{} and PAHs.  Model D
      decreases the uniform total hydrogen density. Models E and F use the
      density profile derived from the model of the \HH{}
      observations~\citep{ha04a,ha04b}. Model F uses a solar gaseous sulfur
      abundance (\ie{} S/H = $10^{-5}$).}
    \label{fig:model} 
  \end{figure*}}

\newcommand{\FigDensity}{%
  \begin{figure}
    \centering %
    \includegraphics[height=\hsize,angle=270]{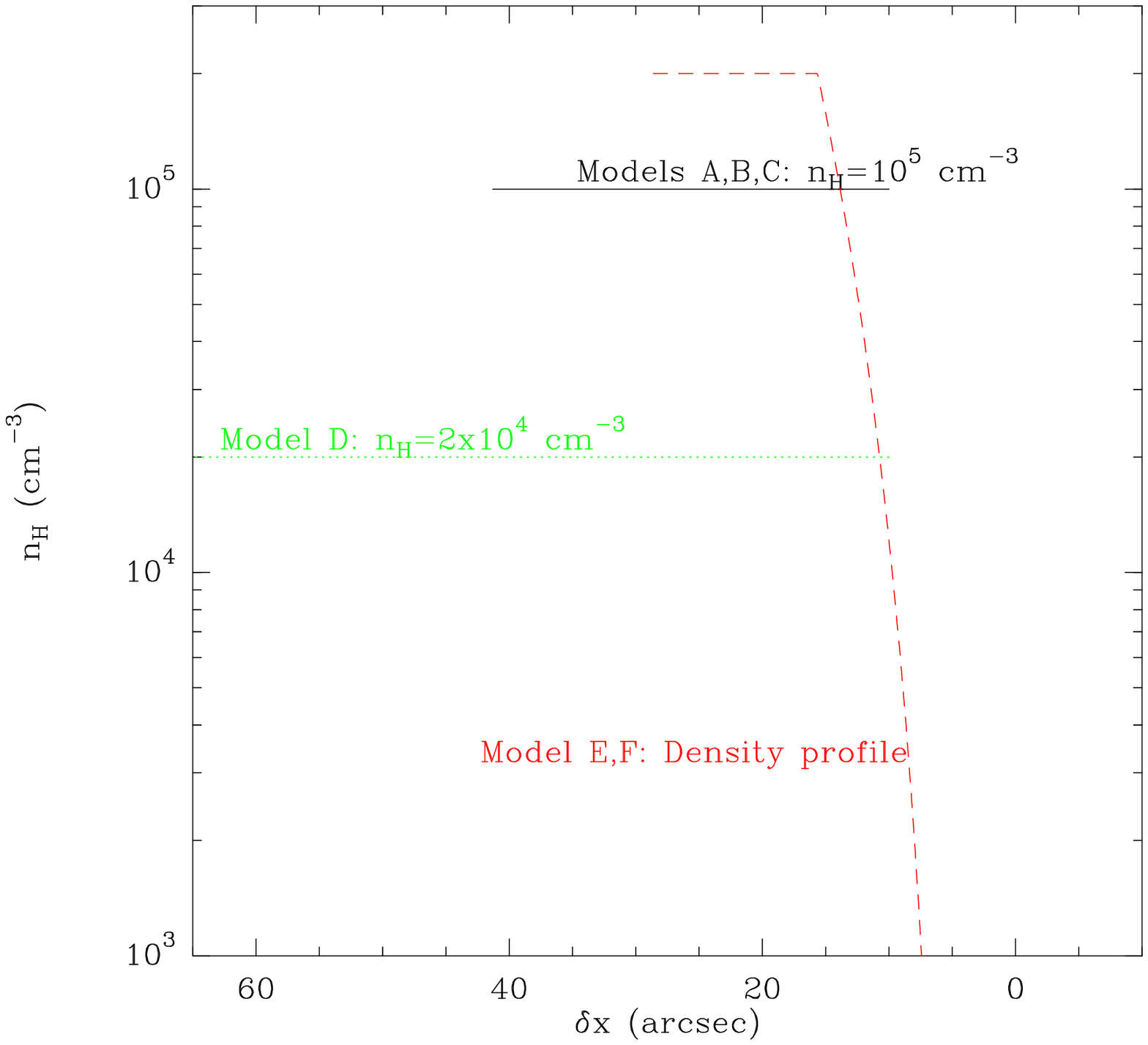}
    \caption{Spatial variation of the total hydrogen density in models A to
      F. In model E and F, the density is increasing as a power law of
      scaling exponent 4 on the first $10''$ and then is kept constant at a
      value of $2\times10^5\pccm$. As in Fig.~\ref{fig:model}, the x-axis
      origin has been set so that \HHexc{} peaks at the position of the
      observed 2.12\mim{}, \HH{} line peak (\ie{} $\delta x =10''$).}
    \label{fig:density}
  \end{figure}}

\newcommand{\FigAbundances}{%
  \begin{figure}
    \centering %
     \includegraphics[height=\hsize{},angle=270]{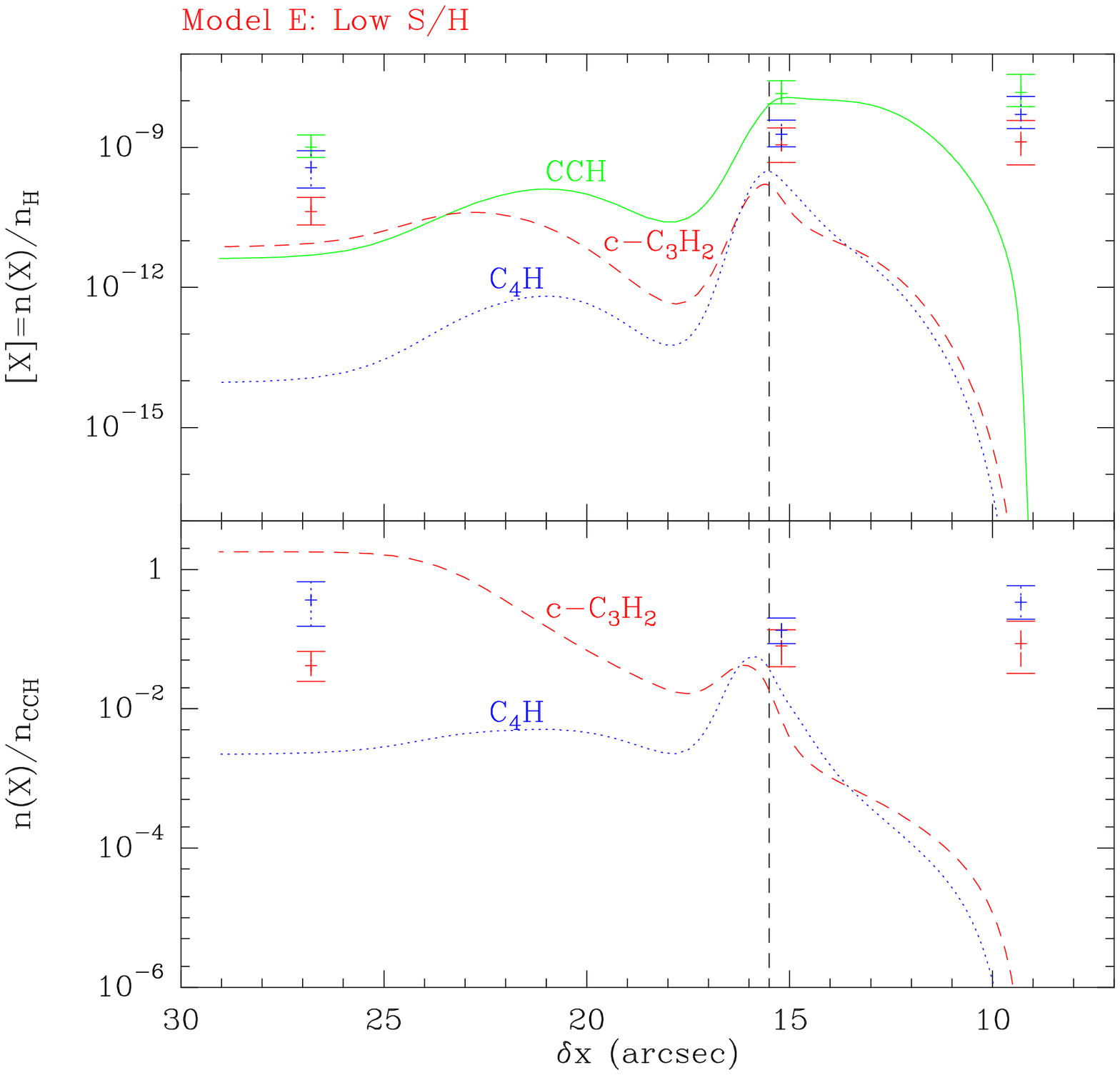}
     \includegraphics[height=\hsize{},angle=270]{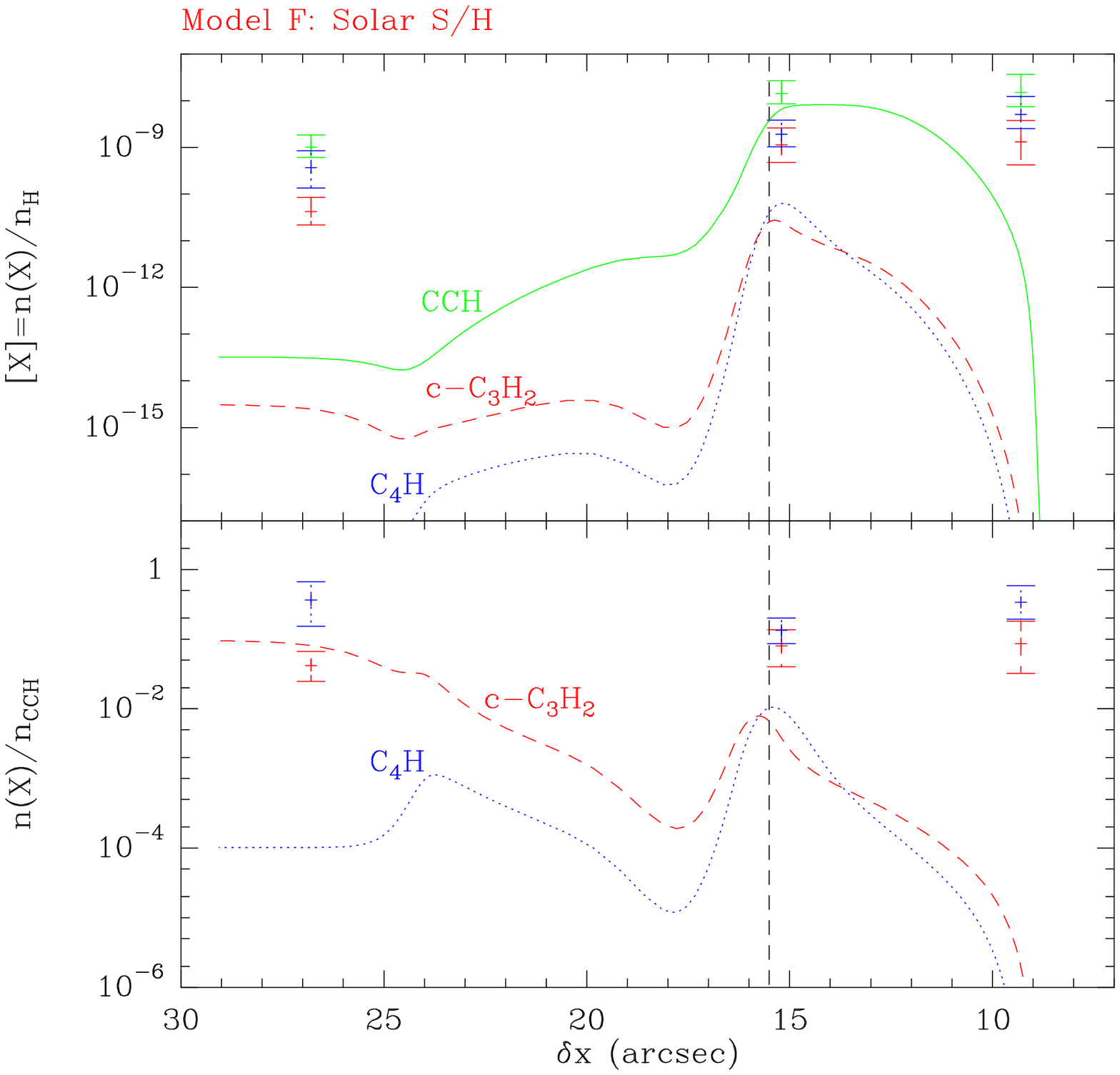}
    \caption{Comparison between our two best models (curves) and 
      observed (points with error bars) abundances of the small
      hydrocarbons. \CCH{} is shown as a green, plain line, \cCCCHH{} as a
      dashed, red line and \CCCCH{} as a blue dotted line.  The top and
      bottom panels respectively show abundances relative to the total
      hydrogen density and \CCH{}.  The dashed vertical line shows the
      position where the total hydrogen density profile swaps from a steep
      gradient to a constant.}
    \label{fig:abundances} 
  \end{figure}}

\begin{document}

\title{Are PAHs precursors of small hydrocarbons in\\
  Photo--Dissociation Regions? The Horsehead case}

\titlerunning{Are PAHs precursors of small hydrocarbons in
  Photo--Dissociation Regions?}

\author{J. Pety \inst{1,2}%
  \and D. Teyssier \inst{3,4}%
  \and D. Foss\'e \inst{1}%
  \and M. Gerin \inst{1}%
  \and E. Roueff \inst{5}%
  \and A. Abergel \inst{6}%
  \and E. Habart \inst{7}%
  \and J. Cernicharo \inst{3}}

\offprints{J. Pety, \email{pety@iram.fr}}

\institute{%
  LERMA, UMR 8112, CNRS, Observatoire de Paris and Ecole Normale
  Sup\'erieure, 24 Rue Lhomond, 75231 Paris cedex 05, France.\\
  \email{fosse@lra.ens.fr, gerin@lra.ens.fr}%
  \and{} IRAM, 300 rue de la Piscine, 38406 Grenoble cedex, France.\\
  \email{pety@iram.fr}%
  \and Instituto de Estructura de la Materia, CSIC, Serrano 121, 28006
  Madrid, Spain.\\
  \email{cerni@damir.iem.csic.es, teyssier@damir.iem.csic.es}%
  \and Space Research Organization Netherlands, P.O.  Box 800, 9700 AV
  Groningen, The Netherlands.%
  \and LUTH UMR 8102, CNRS and Observatoire de Paris, Place J. Janssen
  92195 Meudon cedex, France.\\
  \email{evelyne.roueff@obspm.fr}%
  \and IAS, Universit\'e Paris-Sud, B\^at. 121, 91405 Orsay, France.\\
  \email{abergel@ias.u-psud.fr}%
  \and Osservatorio Astrofisico di Arcetri, L.go E. Fermi, 5, 50125
  Firenze, Italy.\\
  \email{habart@arcetri.astro.it}}

\date{Received 2004; accepted 2005}

\abstract{We present maps at high spatial and spectral resolution in
  emission lines of \CCH{}, \cCCCHH{}, \CCCCH{}, \twCO{} and \CeiO{} of the
  edge of the Horsehead nebula obtained with the Plateau de Bure
  Interferometer (\PdBI{}). The edge of the Horsehead nebula is a
  one-dimensional Photo--Dissociation Region (PDR) viewed almost edge-on.
  All hydrocarbons are detected at high signal--to--noise ratio in the PDR
  where intense emission is seen both in the \HH{} ro-vibrational lines and
  in the PAH mid--infrared bands. \CeiO{} peaks farther away from the cloud
  edge.  Our observations demonstrate that \CCH{}, \cCCCHH{} and \CCCCH{}
  are present in UV--irradiated molecular gas, with abundances
  nearly as high as in dense, well shielded molecular cores.\\
  PDR models \emph{i)} need a large density gradient at the PDR edge to
  correctly reproduce the offset between the hydrocarbons and \HH{} peaks
  and \emph{ii)} fail to reproduce the hydrocarbon abundances. We propose
  that a new formation path of carbon chains, in addition to gas phase
  chemistry, should be considered in PDRs: because of intense
  UV--irradiation, large aromatic molecules and small carbon grains may
  fragment and feed the interstellar medium with small carbon clusters and
  molecules in significant amount. %
  } %
\maketitle
\section{Introduction}
\label{sec:introduction}

Due to the ISO mission, the knowledge of interstellar dust has
significantly progressed in the past years. With its sensitive instruments
in the mid--infrared, ISO revealed the spatial distribution and line
profile of the Aromatic Infrared Bands (AIBs at 3.3, 6.2, 7.7, 8.6 and 11.3
$\mu m$ features), which have shed light on the emission mechanism and
their possible carriers~\citep{bcj00,rjb04}. However, no definite
identification of individual species has been possible yet because the
bands are not specific for individual molecules. The most likely carriers
are large polycyclic aromatic hydrocarbons (PAHs) with about 50 carbon
atoms~\citep{als96,lp03}. The ubiquity of the aromatic band emission in the
interstellar medium has triggered a wealth of theoretical and laboratory
work in the past two decades, which have led to a revision of astrophysical
models. PAHs are now suspected to play a major role in both the
interstellar medium physics and chemistry. With their small size, they are
the most efficient particles for the photo--electric
effect~\citep{bt94,wd01,hvb01}. Their presence also affects the ionization
balance~\citep{fp03,w03}, and possibly the formation of
\HH{}~\citep{hbv04}. The role of PAHs in the neutralization of atomic ions
in the diffuse interstellar medium has been recently reconsidered by
\cite{l03}, following previous work by \citet{lepp88}.  As emphasized soon
after their discovery~\citep{omont86,ld88}, PAHs also play a role in the
gas chemistry: a few laboratory experiments, and theoretical calculations,
suggest that PAHs may fragment into small carbon clusters and molecules
under photon impact (C$_2$, C$_3$, C$_2$H$_2$, etc.)
\citep{job03,lp03,als96,als96b,lhb89,s97}. In addition, investigation of
the lifetimes of interstellar PAHs implies that photo--dissociation may be
the main limiting process for their life in the interstellar
medium~\citep{v01}.

It is therefore appropriate to wonder whether PAHs could fragment
continuously and feed the interstellar medium with small hydrocarbons and
carbon clusters.  This hypothesis is attractive for the following reasons:
\begin{itemize}
\item[\emph{i)}] Cyclopropenylidene (\cCCCHH{}) is widely distributed in
  the interstellar medium~\citep{mi85,mi86,cox88,ll00}.
\item[\emph{ii)}] Recent works have shown that the diffuse interstellar
  medium is more chemically active than previously thought with molecules
  as large as C$_3$~\citep{goi04,o03,abm03,rou02,mai01} and
  \cCCCHH{}~\citep{ll00} widely distributed.  The abundance of C$_3$ and
  \cCCCHH{} are tightly connected with those of smaller molecules, C$_2$
  and \CCH{} respectively, with abundance ratios of [C$_2$]/[C$_3$] $\sim
  10-40$~\citep{o03} and [\CCH{}]/[\cCCCHH{}] $\sim 27.7 \pm
  8$~\citep{ll00}.
\item[\emph{iii)}] \citet{to03} have found a correlation between the
  abundance of C$_2$ and the strength of some (weak) Diffuse Interstellar
  Bands (DIBs).
\end{itemize}
As PAHs are present in the diffuse interstellar medium, could they
contribute in forming both the small carbon clusters (C$_2$, C$_3$) and
larger hydrocarbon molecules which could be the DIBs carriers?

Unfortunately, studies of the PAHs emission bands in the diffuse
interstellar clouds where the carbon clusters have been detected is
extremely difficult because of the low column densities, and also because
the bright background star used for visible spectroscopy prohibits the use
of sensitive IR cameras which would be saturated.  Photo--Dissociation
regions (PDRs) are the first interstellar sources in which the AIBs have
been found and the PAHs hypothesis proposed~\citep{se84,lp84}. It is
therefore interesting to investigate the carbon chemistry in these sources.
\citet{f01} and \citet{tf04} have discussed medium spatial resolution
($30''$) observations of various hydrocarbons in nearby PDRs.  They show
that \CCH{}, \cCCCHH{} and \CCCCH{} are ubiquitous in these regions, with
abundances almost as high as in dark, well shielded, clouds, despite the
strong UV--radiation.  \citet{f03} also report high abundances of \cCCCHH{}
in NGC~7023 and the Orion Bar. Heavier molecules may be present in PDRs as
\citet{tf04} report a tentative detection of C$_6$H in the Horsehead
nebula.  PDRs and diffuse clouds therefore seem to share the same carbon
chemistry, but because of their larger \HH{} column density and gas
density, PDRs offer more opportunities for detecting rare species.

\citet{tf04} and \citet{f03} reckon the presence of carbon chains is in
favor of a causal link between small hydrocarbons and PAHs, but they lack
the spatial resolution to conclude. In the present work, we present high
spatial resolution observations of one source studied by \citet{tf04}, the
Horsehead nebula, obtained with the Plateau de Bure interferometer. We
describe the observations in Sec.~\ref{sec:obs}. We show the interferometer
maps in Sec.~\ref{sec:results}. Finally, Sec.~\ref{sec:discussion} presents
a comparison with chemical models.

\section{Observations and data reduction}
\label{sec:obs}

\subsection{The Horsehead nebula}

The Horsehead nebula, also called Barnard 33, appears as a dark patch of
$\sim 5'$ extent against the bright HII region IC434. Emission from the gas
and dust associated with this globule has been detected from mid--IR to
millimeter wavelengths~\citep{ab02,ab03,tf04,prb03}.  From the analysis of
the ISOCAM images, \citet{ab03} conclude that the Horsehead nebula is a PDR
viewed edge-on and illuminated by the O9.5V star $\sigma$Ori at a projected
distance of 0.5\deg{} (3.5\pc{} for a distance of 400\pc{}, \citet{at82}).
The far--UV intensity of the incident radiation field is $G_0 = 60$
relative to the average interstellar radiation field in Draine
units~\citep{dra78}. The gas density, derived from the excitation of
molecular lines, and from the penetration depth of the UV--radiation, is a
few $10^4\pccm$~\citep{ab03}.  From a combined analysis of maps of both CO
and atomic carbon, \citet{lchamp} obtain similar figures for the gas
density.  \citet{ha04a,ha04b} have modeled the emission of \HH{} (from
narrow band images of \HH{} ro-vibrational line), PAHs and CO, and conclude
that \emph{i)} the gas density follows a steep gradient at the cloud edge,
rising to $n_\emr{H} = 10^5\pccm$ in less than $10''$ (\ie{} 0.02\pc{}) and
\emph{ii)} this density gradient model is essentially a constant pressure
model (with $P = 4 \times 10^6\Kkms$).

The edge of the Horsehead nebula is particularly well delineated by the
mid--IR emission due to PAHs, with a bright 7.7\mim{}--peak (hereafter
named ``IR peak'') reaching 25~MJy/sr at $\alpha_{2000} = 05^h40^m53.70^s ,
\delta_{2000} = -02^\circ 28' 04''$.  Fig.~\ref{fig:mosaic} shows the
region observed with the \IRAM{} \PdBI{} centered near the ``IR peak''.
Two mosaics (one for hydrocarbon lines and the other for the CO lines) have
been observed. Their set-ups are detailed in Table~\ref{tab:pdb}.

\TabObs{} %
\FigMosaic{} %

\subsection{Observations}

\subsubsection{\cCCCHH{} and \CCCCH{}}

First \PdBI{} observations dedicated to this project were carried out with
6 antennas in CD configuration (baseline lengths from 24 to 229~m) during
March--April 2002. The 580\MHz{} instantaneous IF--bandwidth allowed us to
observe simultaneously \cCCCHH{} and \CCCCH{} at 3\mm{} using 3 different
20\MHz{}--wide correlator windows. One other window was centered on the
\CeiO{} (\Jtwo{}) frequency. The full IF bandwidth was also covered by
continuum windows both at 3.4 and 1.4\mm{}.  \cCCCHH{} and \CCCCH{} were
detected but the weather quality precluded use of 1.4\mm{} data.

We observed a seven--field mosaic on a compact, hexagonal pattern, with
full Nyquist sampling at 1.4~mm and large oversampling at 3.4~mm. This
mosaic, centered on the IR peak, was observed during about 6h on--source
observing time per configuration. The rms phase noises were between 15 and
40\deg{} at 3.4\mm{}, which introduced position errors $< 0.5''$. Typical
3.4\mm{} resolution was $6''$.

\subsubsection{\CCH{} and \CeiO{}}

\TabFlux{} %

As a follow-up, we carried out observations of \CCH{} at \PdBI{} with 6
antennas in CD configuration during December 2002 and March 2003. We used a
similar correlator setup: three 20\MHz{}--wide windows were centered so as
to get the four 3.4\mm{} hyperfine components of \CCH{}; one 20\MHz{}--wide
window was centered on the \CeiO{} (\Jtwo{}) frequency; the remaining
windows were used to observe continuum at 3.4 and 1.4\mm{}.

Exactly the same mosaic (center and field--pattern) and approximately the
same on--source observing time per configuration ($\sim$ 6h) as before were
used.  The rms phase noises were between 10 and 40\deg{} except during 4
hours in D configuration where they were between 8 and 20\deg{} at
3.4\mm{}. The data of those 4 hours have been used to build the \CeiO{} map
as the 1\mm{} phase noises were then low enough (between 20 to 45\deg{}).
We thus ended up with a $6''$ typical resolution both at 3.4\mm{} and
1.4\mm{}. Both \CCH{} and \CeiO{} were easily mapped while no continuum was
detected at a level of 2~mJy/beam in a $6''$--beam.

\subsubsection{\twCO{}}

As part of another project (A.~Abergel, private communication), the \twCO{}
(\Jone{}) and \twCO{} (\Jtwo{}) lines were simultaneously observed during
6h on--source at \PdBI{} in November 1999 (only 5 antennas were then
available) in configuration C (baseline lengths from 24 to 82~m).
\FigMaps{} %
\FigMapsRot{} %
The observation consisted in a 4--fields mosaic, fully sampled at 1.3\mm{}.
The mosaic center is slightly shifted compared to the two other
observations.  Weather was excellent with phase noises from 3 to 5\deg{}
and 6 to 10\deg{} at 2.6\mm{} and 1.3\mm{}, respectively. Typical
resolutions were $5''$ at 2.6\mm{} and $2.5''$ at 1.3\mm{}.

\subsubsection{Other data: \HH{}, ISO-LW2 and 1.2\mm{} dust continuum}

The \HH{} v=1-0 S(1) map shown here is a small part of Horsehead
observations obtained at the NTT using SOFI. The resolution is $\sim 1''$.
Extensive explanations of data reduction and analysis are discussed
elsewhere~\citep{ha04a,ha04b}. The ISO-LW2 map (published by~\citet{ab03})
shows aromatic features at $7.7 \mu m$ with a resolution of $\sim 6''$. The
1.2\mm{} dust continuum has been obtained at the \IRAMthm{} telescope with
a resolution of $\sim 11''$ and has already been presented by~\citet{tf04}.

\subsection{\PdBI{} data processing}

All data reduction was done with the \GILDAS{}\footnote{See
  \texttt{http://www.iram.fr/IRAMFR/GILDAS} for more information about the
  \GILDAS{} softwares.} softwares supported at \IRAM{}. Standard
calibration methods using close calibrators were applied to all the \PdBI{}
data. The calibrator fluxes used for the absolute flux calibration are
summarized in Table~\ref{tab:fluxes}.

Following~\citet{gu96}, single--dish, fully sampled maps obtained at the
\IRAMthm{} telescope~\citep{tf04,ab03}) were used to produce the
short--spacing visibilities filtered out by every mm-interferometer (\eg{}
spatial frequencies between 0 and 15\m{} for \PdBI{}). Those
pseudo-visibilities were merged with the observed, interferometric ones.
Each mosaic field were then imaged and a dirty mosaic was built combining
those fields in the following optimal way in terms of signal--to--noise
ratio~\citep{gu01}:
\begin{displaymath}
\displaystyle %
    J(\alpha,\delta) = \sum\nolimits_i \frac{B_i(\alpha,\delta)}{\sigma_i^2}\,F_i(\alpha,\delta)
    \left/
      \displaystyle \sum\nolimits_i \frac{B_i(\alpha,\delta)^2}{\sigma_i^2}.
    \right.  
\end{displaymath}
In this equation, $J(\alpha,\delta)$ is the brightness distribution in the
dirty mosaic image, the $B_i$ are the response functions of the primary
antenna beams, the $F_i$ are the brightness distributions of the individual
dirty maps, and the $\sigma_i$ are the corresponding noise values. As may
be seen on this equation, the dirty intensity distribution is corrected for
primary beam attenuation. This implies that noise is inhomogeneous. In
particular, noise strongly increases near the edges of the field of view.
To limit this effect, both the primary beams used in the above formula and
the resulting dirty mosaics are truncated.  Standard level of truncation is
set to 20\% of the maximum in \GILDAS{}. In our case, the intensity
distribution does not drop to zero at all field edges.  Hence, we used a
much lower level of truncation of the beam (\ie{} 5\%) to ensure a better
deconvolution of the side lobes of the sources sitting just at the field
edges. We then use the standard adaptation to mosaics of the H\"ogbom CLEAN
algorithm to deconvolve~\citep{gu01}. The sharp edge of the \HH{} emission
defines a boundary that may be used as an \emph{a priori} knowledge in the
deconvolution of the \PdBI{} images: We use this boundary as a numerical
support (in the language of signal processing) to forbid the search of
CLEAN components outside the PDR front (\ie{} in the direction of the
exciting star).  We finally truncated the noisy clean mosaic edges using
the standard truncation level. It may be noted that the \CCCCH{} maps are
particularly difficult to deconvolve due to their low signal--to--noise
ratio, $S/N < 10$ to 15.

\section{Results}
\label{sec:results}

\FigCorrel{} %
\FigCuts{} %

\subsection{Maps}

The \PdBI{} maps are shown in Fig.~\ref{fig:maps1} and~\ref{fig:maps2}
together with the 7\mim{} ISOCAM image~\citep{ab03}, the 1.2\mm{} dust
emission map~\citep{tf04} and the map of the \HH{} 2.1\mim{} line
emission~\citep{ha04a,ha04b} for comparison. For all lines, we obtained
excellent spatial resolutions, similar to or even better than the ISOCAM
pixel size of $6''$ (see Table \ref{tab:pdb}). Fig.~\ref{fig:maps1} shows
the maps in the natural Equatorial coordinate system while
Fig.~\ref{fig:maps2} shows the maps in a coordinate system where the x-axis
is in the direction of the exciting star and the y-axis defines an
empirical PDR edge that corresponds to the sharp boundary of the \HH{}
emission (\ie{} the maps have been rotated by 14\deg{} counter-clockwise
and horizontally shifted by $20''$). The latter presentation enables a much
better comparison of the PDR stratification.

The main structure in all hydrocarbon maps is an approximatively N-S
filament, following nicely the cloud edge and corresponding closely to the
mid--IR filament on the ISO-LW2 image. A weaker and more extended emission
is also detected, which has no counterpart in the ISO-LW2 image and can be
attributed to the bulk of the cloud. It is interesting to note that the
hydrocarbon emission presents a minimum behind the main filament, and a
weaker secondary maximum within the extended emission. The hydrocarbon
emission is stronger on the edges of the dust 1.2\mm{} emission and avoids
the region of maximum dust emission where the gas is likely denser. This
confirms a tendency revealed by chemical surveys of dense cores (study of
TMC-1 by \citet{pratap97} and L134N by \citet{dickens00,fosse}): \ie{}
carbon chains (\CCH, \CCCCH,...)  generally avoid the densest and more
depleted cores.

Even at the high spatial resolution provided by the plateau de Bure
Interferometer, the maps of all hydrocarbons remain very similar.  Detailed
inspection of the maps shows small differences between \CCH{} and
\cCCCHH{}, but these do not affect the overall similarity. Indeed, the
joint histogram describing the correlation of line maps for \emph{i)} the
two most intense \CCH{} lines, \emph{ii)} \cCCCHH{} and \CCH{}, and
\emph{iii)} \CCCCH{} and \CCH{} are displayed in Fig.~\ref{fig:correl}. As
expected the two \CCH{} lines are extremely well correlated as illustrated
by the elongated shape (approaching a straight line) of the joint
histogram. The correlations between \cCCCHH{} and \CCH{}, and between
\CCCCH{} and \CCH{} are excellent too, though the signal--to--noise ratio
is not as good for \CCCCH{}.  For this plot, we have used all points lying
inside the support used for the deconvolution.

The high resolution \cCCCHH{} map appears to show more structure than the
\CCH{} maps, particularly in the well shielded cloud interior (on the left
hand side of the main filament). This effect seems real since it does not
appear for the satellite \CCH{} line maps, which have similar intensities
and signal--to--noise ratio as the \cCCCHH{} map. The \CCCCH{} maps are too
noisy for a detailed analysis but are nevertheless very well correlated
with the \CCH{} map.  {\em The correlations found at low spatial
  resolution~\citep{tf04} are not an artifact but persist at high spatial
  resolution.}

The correspondence of hydrocarbons with CO and \CeiO{} is not as good. The
\CeiO{} (\Jtwo{}) map presents two maxima, located on either side of the
\CCH{} peak along a N-S direction: The \CCH{} peak is associated with a
local minimum of \CeiO{} emission. Also, the \CeiO{} emission peak is
displaced farther inside the cloud (East) compared to \CCH{} and the other
hydrocarbons.

To illustrate further the differences in the spatial distribution of CO,
\CeiO{} and the hydrocarbons, we show two series of cuts across the PDR in
Fig.~\ref{fig:cuts}.  The UV--radiation comes from $\sigma$ Ori far to the
right side of the Fig.~\ref{fig:maps2}. The cuts have been taken along the
$\sigma$ Ori direction (\ie{} PA$=-104\deg{}$). The main peak for all
hydrocarbons is located near an offset of $\delta x \simeq 12-15\arcsec$ at
less than $5''$ of the \HH{} peak. The ISO-LW2 peak is located half--way
between hydrocarbons and \HH{} peaks. Intense \twCO{} emission, in both the
\Jone{} and \Jtwo{} lines is also detected in the same region, while the
\CeiO{} (\Jtwo{}) emission rises farther (at least $5''$) inside the cloud.

\FigCOAnalysis{} %
\FigCOSpectra{}  %

As shown in Fig.~\ref{fig:co:analysis}, the \twCO{} (\Jtwo{}) emission
(convolved at the same angular resolution as the \twCO{} \Jone{}
transition) is very bright ($\geq 50\K{}$ at $10.6\kms$, the line peak
velocity) and more intense than \twCO{} (\Jone{}) in the most external
layers of the PDRs, facing directly $\sigma$Ori. The line intensity ratio
$T_b(1-0)/T_b(2-1)$ rises from $\sim 0.3$ to $\sim 0.8$ from West to East.
Combined with the high brightness temperature detected for both lines, the
higher brightness temperature of the \twCO{}(2-1) line is a clear sign of
the presence of warm and dense gas.  We have estimated the kinetic
temperature using an LVG model. We assumed that the emission is resolved
and fills the beam. We explored the kinetic temperature dependence upon the
density by solving for 5 different proton densities going from
$1.6\times10^4\pccm$ to $10^5\pccm$. Under these hypotheses, the \twCO{}
line intensity ratio and brightness temperature constrain the kinetic
temperature to increase from $60\K$ in the inner PDR ($15''= 0.03\pc$ from
the PDR edge) to more than $100\K$ in the outer layers for proton densities
larger than $4\times10^4\pccm$. For lower proton densities, the kinetic
temperature still starts from $60\K$ in the inner PDR but increases much
more stiffly.  The kinetic temperature derived from single dish
observations~\citep{ab03} is lower, in the $30-40\K$ range and corresponds
to the bulk of the cloud, rather than to the warm UV--illuminated edge.

\subsection{Abundances}

We have computed the CO and hydrocarbons column densities at three
representative positions in the maps: the ``IR peak'' where the PAH and
hydrocarbon emission is the largest, the ``IR edge'' $10''$ West which
represents the region with the most intense UV--radiation and a ``Cloud''
position behind the IR filament. Table~\ref{tab:abondances} lists the
derived column densities and abundances relative to the total number of
protons for these 3 positions.  We have used a LVG model with different
uniform total hydrogen density\footnote{``Total hydrogen density'' is a
  short cut for the total density of hydrogen in all forms.} (from
$10^4\pccm$ to $10^5\pccm$) and a kinetic temperature of 40\K{} for the
``cloud'' position, and between 60 and 100\K{} for the IR positions.  The
variance of the column densities therefore reflects both the systematic
effect due to the imperfect knowledge of the physical conditions, and the
random noise of the data. In most cases, the former contribution is the
largest. The \HH{} column densities are derived from the dust 1.2\mm{}
emission assuming the same dust properties for all positions but a dust
temperature range of 20 to 40\K{} for the ``Cloud'' position and 40 to
80\K{} for the IR positions.

\TabAbundances{} %

The LVG solution implies a typical \twCO{} column density of $2 \times
10^{17}\pscm$. This is inconsistent with the derived column density of
\CeiO{} and the local ISM $^{16}$O/$^{18}$O elemental
ratio~\citep[560,][]{wr94}.  Fig.~\ref{fig:co:spectra} shows clear
indications of self-absorption of the \twCO{} spectra (asymmetries and dips
in the top of the line profiles) while the \CeiO{} spectra are Gaussian.
The same behaviour is seen in the single dish data discussed
by~\citet{ab03} (cf.\ their Fig.~5).  This explains why the LVG solution
does not succeed in correctly inferring the \twCO{} column density.
Conversely, the \CeiO{} abundance relative to H is fairly constant for all
positions at $[\CeiO] = 1.0 \times 10^{-7}$.  Assuming a local ISM
$^{16}$O/$^{18}$O elemental ratio, this corresponds to a CO abundance
relative to the total number of hydrogen atoms of $[\mbox{CO}] = 5.6 \times
10^{-5}$, in rather good agreement with the gas phase abundance of carbon
derived from CO in warm molecular clouds, and to the carbon abundance in
diffuse clouds~\citep{lk94,sm01}.  In addition, using IRAM-30m spectra of
\thCO{} and \CeiO{} published by~\citet{ab03}, we found $[\thCO]/[\CeiO]
\sim 7$. This good agreement with the local ISM isotopic ratio make us
confident that we can use our LVG analysis on the PdBI \CeiO{} spectra to
estimate the CO density. According to~\citet{lchamp}, atomic carbon is less
abundant than CO in the PDR. The peak column density of neutral carbon,
observed with a $15''$ beam, is $\sim 1.6 \times 10^{17}$ cm$^{-2}$
corresponding to a carbon abundance of $[\mbox{C}] = 5 \times 10^{-6}$.
Even if we take into account the difference in linear resolution, we don't
expect an increase of the column density larger than a factor of two based
on the comparison of the low resolution single dish data with the
interferometer maps of other tracers.  Finally, though the \HH{} column
densities are fairly similar at the ``IR peak'' and ``cloud'' positions,
the abundances of hydrocarbons are larger by a factor of at least 5.0 at
the ``IR peak''.  The abundances seem to be even larger at the ``IR edge''
than at the cloud position.

\section{Discussion}
\label{sec:discussion}

\subsection{Comparison with models}

\FigModel{} %

We have used a monodimensional PDR
code~\citep[\texttt{http://aristote.obspm.fr/MIS/}]{lep02} for modeling the
observations of the Horsehead nebula. The slab geometry is locally
appropriate as seen on Fig.~\ref{fig:maps2}. We did not take into account
projection effects as the source is viewed almost edge-on. Indeed,
\cite{ha04b} show that the main effect of the PDR possible small
inclination ($< 6\deg$) is to enlarge the peak profiles and to shift them
all compared to the model edge. The model includes a detailed treatment of
the photo--dissociation of \HH{} and the CO isotopes as well as the
statistical equilibrium of their rovibrational (rotational, respectively)
states in a steady state approach. The parameters of the model include the
elemental abundances, the cosmic ray ionization rate, the scaling factor of
the interstellar ultraviolet radiation field (ISRF) measured in Draine
units, the density profile and the grain parameters. As the observations
involve complex carbon molecules, we have used the so called ``new standard
model'' chemical rate file of Herbst and collaborators~\citep{lrp98},
available on the web
site\footnote{\texttt{http://www.physics.ohio-state.edu/\~{}eric/research\_}
  \texttt{files/cddata.july03}}.  In a previous paper~\citep{tf04}, we have
found that the other extensive chemical rate file provided by the UMIST
group~\citep{ltm00} gave close results on the carbon chain molecules.  As
\CeiO{} observations are reported, we have added to this reaction set, the
main isotopic molecules involving $^{18}$O and introduced the corresponding
fractionation reactions~\citep{gra82}. We have also introduced the
photo--dissociation rates given by~\citet{evd88}, when available, which
have been calculated specifically with the Draine ISRF and which were
different from the values reported in the chemical rate file.  The
resulting chemical network involves about 450 chemical species and 5000
reactions. Only the most stable isomeric forms of hydrocarbons are
considered here.

We define a reference model (hereafter named model A) of the Horsehead
nebula as a uniform sheet of gas and dust of total hydrogen density
$n_\emr{H} = 10^5\pccm$ exposed to a ISRF of 100 measured in Draine units.
The cosmic ray ionization rate has a value of
$5\times10^{-17}~\mbox{s}^{-1}$ and the elemental abundances are as
follows: C/H = $1.38\times10^{-4}$, O/H = $3.02\times10^{-4}$, $^{18}$O/H =
$6\times10^{-7}$, N/H = $7.95\times10^{-5}$, S/H = $5.8\times10^{-8}$, Cl/H
= $1.86\times10^{-9}$, P/H = $9.3\times10^{-10}$, Fe/H =
$1.7\times10^{-9}$, Mg/H = $10^{-8}$, Na/H = $2.3\times10^{-9}$. The
properties of the grains are the same as described in~\citet{lep02}, \ie{}
the size distribution law is taken from~\cite{mat77} with an exponent of
$-3.5$ and we describe the attenuation of grains from the far--UV to the
visible via the galactic extinction curve given as an analytic function of
1/$\lambda$ by including the coefficients derived by~\cite{fm88}. Charge
exchange reactions between \Cp{} and PAHs are not taken into account. The
gas to dust mass ratio is 100.

Fig.~\ref{fig:model} shows \emph{i)} the abundance of the \HH{}
rovibrationnally excited in the v=1, J=3 level at the origin of the
2.12\mim{} line (this abundance is hereafter referred to as \HHexc{}) and
\emph{ii)} the C, CO and hydrocarbon abundances for this reference model
and 5 variants. We ensure that the \HHexc{} peak position is set at $\delta
x = 10''$ as in the observations. Our reference model correctly reproduces
the observed 3 to $5''$ offset between the hydrocarbon and \HH{} peaks.
The \CeiO{} also peaks behind the hydrocarbons at $\delta x = 20-25''$.
However, the \HH{} profile is not correctly modeled here.

In model B, we replaced the Galactic extinction curve by one more
representative of molecular gas. We have chosen HD~147889 in Ophiuchus.
Its extinction curve has a rather strong far--UV rise ($E_{B-V} =
1.09$~\citet{fm88}). Its ratio between the total and selective extinctions,
\Rv{}, is 4.2 a typical figure for molecular gas~\citep{gc03,ccm89}. The
PDR stratification does not qualitatively change compared to model A: It is
just compressed. In model C, we added reactions of charge exchange between
\Cp{} and PAHs. This enhances the neutral atomic carbon abundance but does
not have a large effect on the hydrocarbons: Only \CCH{} peaks closer from
the \HH{} peak compared to model A. Neither model B nor C improves the
modeling of the \HH{} profile.

\FigDensity{} %

As shown by model D, E and F, the density structure has a major impact on
the PDR structure. Fig.~\ref{fig:density} shows the density profiles
associated to each model. When keeping the total hydrogen density uniform
but decreasing its value to $2 \times 10^4\pccm$ (as in model D), the
carbon and hydrocarbon abundance peaks highly broadened and shifted inward
by more than $20''$, a prediction clearly violated by the high resolution
\PdBI{} data. Model E and F uses a density profile provided
by~\citet{ha04a,ha04b} to fit the 2.12\mim{}--\HH{} emission. Indeed, the
\HHexc{} profile qualitatively changes (it is now a peak rising from zero
at the PDR edge) but it also reproduces the \HH{} filament width. Those two
models which impose a steep total hydrogen density gradient at the PDR
edge, are the only ones that succeed to correctly reproduce the offset
between the hydrocarbon and \HH{} peaks as well as the form of the \HH{}
peak. The sole difference between models E and F is the gaseous sulfur
abundance: sulfur is depleted from the gas phase in model E (S/H = $5.8
\times 10^{-8}$) while the gaseous sulfur abundance is solar in model F
(S/H = $10^{-5}$).

\FigAbundances{} %

Fig.~\ref{fig:abundances} is a zoom for our two best models (\ie{} E and F)
of the spatial variations of the abundances of hydrocarbons relative to
\emph{i)} total hydrogen density (top panel) and \emph{ii)} \CCH{} (bottom
panel).  The observed abundances are overplotted with their error bars.
The dashed vertical line separates the zone where the proton gas density is
constant from the zone where the proton gas density rapidly decreases
outward. This latter zone is associated with the PDR.  The sulfur elemental
abundance has different effects in those two regions. In the region of
moderate visual extinction (\ie{} the``IR edge'' and the ``IR peak'' where
$\Av{} \la 1$), the charge transfer reaction between C$^+$ and S leading to
S$^+$ and C reinforces the abundance of neutral carbon and thus enables the
formation of carbon chains via the rapid neutral-carbon atom reactions.
However this effect is small. Indeed this is in the dark region where the
sulfur elemental abundance has a large effect. When the sulfur abundance is
solar, the small carbon chains $\rm C_2$, CCH, $\rm C_2H_2$, $\rm C_3H$,
$\rm C_3H_2$ and $\rm C_4H$ react with $\rm S^+$ to give $\rm C_2S^+$, $\rm
CCS^+$, $\rm HC_2S^+$, $\rm C_3S^+$, $\rm HC_3S^+$ and $\rm C_4S^+$. In
this main destruction path of the small carbon chains, one hydrogen atom is
released impairing the reformation of the carbon chains. When S is higly
depleted as in Model E, this destruction mechanisms is superseded by other
paths involving $C^+$. Those paths form carbon chain ions which in turn
contributes to the formation of other carbon chains.  Overall, model E
(\ie{} low S/H) performs better in the comparison with observed abundances.
The only exception is the $n(\cCCCHH)/n(\CCH)$ ratio at the ``cloud''
position. We will thus use model E only for comparison with the
observations. At the IR peak (median point at $\delta x = 15.5''$), \CCH{}
abundance is correctly reproduced while \cCCCHH{} and \CCCCH{} abundances
are underestimated by at least a factor of 3. Discrepancies are much higher
both at the ``cloud'' (point to the right at $\delta x = 27''$) and the
``IR edge'' (point to the left at $\delta x = 9''$) positions.  In the
UV--illuminated edge, the modeled $[\CCH{}]$ has a quite shallow increase
with $\delta x$ while the modeled $[\cCCCHH]$ and $[\CCCCH]$ share the same
steep abundance profile.  By contrast, the observed (``IR edge'')
abundances are very similar for the 3 species, reflecting the very good
spatial correlation between the different hydrocarbons (see
Fig.~\ref{fig:correl}).  This discrepancy is independent of our knowledge
of the total hydrogen density as it is also seen when comparing abundances
relative to \CCH{}.

In summary, \emph{none} of our models is able to correctly reproduce the
relative stratification of \HH{} and small hydrocarbons. Comparison of
model A and C shows that to reproduce the observed offset between
hydrocarbon and \HH{} peaks, we need a high total hydrogen density
($10^5\pccm$). By varying the profile density (model E and F), a shallow
total hydrogen density increase at the PDR edge is needed for reproducing
the profile of the 2.12\mim{} \HH{} line. However, the shallower the total
hydrogen density increase, the larger the modeled offset between \HH{} and
the hydrocarbons. A good compromise is provided by a total hydrogen density
profile increasing as a power law with a scaling exponent 4 on the first
$10''$ and then constant at a value of $2\times10^5\pccm$.  \citet{ha04b}
shows that this model essentially correspond to a constant pressure model
(with $P = 4 \times 10^6\Kkms$). Finally, model E with low sulfur elemental
abundance performs better than model F with solar abundance. Nonetheless,
even model E do not succeed to reproduce the good hydrocarbon correlation
seen in the illuminated part of the PDR: while \CCH{} is correctly
predicted to have a smooth abundance increase, modeled \cCCCHH{} and
\CCCCH{} abundances show a much too steep increase.

\subsection{Can the fragmentation of PAHs contribute to the synthesis
  of small hydrocarbons?}

Examining the model predictions in more details, three hypotheses can be
proposed for explaining the discrepancies between model calculations and
observations:
\begin{itemize}
\item[\emph{i)}] The photo--dissociation rates used in the models may be
  incorrect.  As the main destruction process near the cloud edge is
  photo--dissociation, the actual values of the photo--dissociation rates
  are critical for an accurate prediction. However, similar results are
  obtained with the UMIST95 and NSM rate files. The photo--dissociation
  rates for \cCCCHH{}, C$_3$H and \CCCCH{} are 10$^{-9}$ s$^{-1}$ for both
  rate files, and differ by a factor of two for \CCH{} (\ie{} $0.51 \times
  10^{-9}$ s$^{-1}$ for UMIST95 and $ 10^{-9}$ s$^{-1}$ for NSM). The
  photo--dissociation rates of larger chains are similar.  In most cases,
  except for \CCH{} and acetylene, the numbers given in the rate files are
  not well documented. For instance, \citet{evd88} discusses the
  photo--dissociation rate of \cCCCHH{} and concludes that it is accurate
  within an order of magnitude. More accurate photo--dissociation rates are
  clearly needed for the carbon chains and cycles. Recent calculations have
  been performed for \CCCCH{} showing that the photo--dissociation
  threshold is 5.74~eV, but that efficient photo--dissociation requires
  more energetic photons, typically above 6.5~eV~\citep{ggl01}. However, it
  is unlikely that the rates are low enough to explain the large
  discrepancies between the models and the data since these molecules are
  known to be sensitive to UV--radiation~\citep{jac91,song94}.
\item[\emph{ii)}] Another possibility is that the chemical networks are
  missing important reactions for the synthesis of hydrocarbons.
  Neutral--neutral reactions are progressively included in the rate files,
  but still are much less numerous than ion--molecule reactions.  It is now
  known that atomic carbon, diatomic carbon and \CCH{} may react with
  hydrocarbons~\citep{kv03,sss02,mk02}. More work remains to be done.
  However, preliminary tests using a more extended data base of chemical
  reactions have not led to significant improvement.
\item[\emph{iii)}] The excellent spatial correlation between the mid--IR
  emission due to PAHs and the distribution of carbon chains suggests a
  last hypothesis: the fragmentation of PAHs due to the intense far
  UV--radiation could seed the interstellar medium with a variety of carbon
  clusters, chains and rings~\citep[and references
  therein]{s97,v01,lp03,job03}.  These species would then further react
  with gas phase species (C, \Cp{}, H, \HH{}, etc.) and participate in the
  synthesis of the observed hydrocarbons. \citet{f03} also favor this
  explanation for explaining the abundance of \cCCCHH{} they observed in
  other PDRs.
\end{itemize}

A correct exploration of this third hypothesis needs a good qualitative and
quantitative description of both the fragmentation and reformation of PAHs,
which is out of the scope of this paper. We here give only a few
indications. \citet{omont86} pioneered the attempts to understand the role
of PAHs in the interstellar chemistry. Elaborating on this work,
\citet{ld88} suggested that participations of PAHs in the ion chemistry of
\emph{dense clouds} lead to large increase in the abundances of small
hydrocarbons. Indeed, when the PAH fractional abundance exceeds $\sim
10^{-7}$, the formation of PAH$^-$ triggers mutual neutralization of the
positive atomic and molecular ions and introduces new pathways for the
formation of complex molecules.  The equilibrium abundances of neutral
atomic carbon C, \CCH{} and \cCCCHH{} may thus be enhanced by two order of
magnitude. By comparison, our model C which includes charge exchange
between \Cp{} and PAHs, shows a decrease of \CCCCH{} and \cCCCHH{}
abundances by at least an order of magnitude in the dark region.
Introduction of mutual neutralization between \Cp{} and PAH$^-$ could be an
interesting alternative to our ``artificial'' lowering of the sulfur
abundance. We are currently acquiring CS data at PdBI to constrain the S
chemistry independently.
  
\citet{lepp88} suggested that the ion chemistry of \emph{diffuse clouds}
has little impact on the CH, OH and HD abundance, but can lead to large
increase in the abundance of other species (\HH{}, NH$_3$ and most
noticeably $\rm CH_4$ and $\rm C_2H_2$) by successive reactions of PAH and
PAH$^-$ with carbon and hydrogen atoms. \citet{tal93} suggested that
Coulombic explosion of doubly ionized PAH could create \cCCCHH{} through
the electronic dissociative recombination of $\rm C_3H_3^+$.  Laboratory
experiments by \citet{jochims94} suggested that PAHs with less than 30-40
carbon atoms will be UV--photodissociated in H\textsc{i} regions while
larger ones will be stable.  Based on those results, models by
\citet{als96,als96b} indicate that only PAHs with more than 50 carbon atoms
survive the high UV radiation field of the diffuse interstellar medium,
whereas smaller PAHs such as coronene or ovalene are destroyed by the loss
of acetylenic groups.  Destruction timescales are a few years for neutral
species and typically five time shorter for the corresponding cations. All
those reactions starts from neutral or cation PAHs. They will be in
competition with charge exchange and mutual neutralization discussed above.
Mutual neutralization has a maximal effect in the transition region where
the gas is molecular but the electronic abundance is significant. This
region corresponds more or less to the region of maximum emission from the
PAHs or slightly deeper in the molecular cloud.  All other cited reactions
are more efficient toward the illuminated edge where PAHs are mainly
neutral.  Recently, \citet{lp03} discussed the possibility of addition
reactions with ionized carbon, starting from the high reaction rate between
\Cp{} and anthracene measured by~\citet{can95}.  If similar reaction rates
persist for heavier PAHs, addition reaction with carbon would be very
efficient for counteracting the destruction by far--UV photons.

From the observational point of view, the mid--IR emission due to PAHs is
extended in interstellar clouds. On the other hand, a detailed analysis of
the mid and far--IR images obtained by IRAS led~\citet{bou90}
and~\citet{Ber93} to conclude that PAHs disappear in the dense cold cloud
interiors, probably because they coagulate and/or condense. \citet{step03}
describe a convincing case for such a process in a small filament of the
Taurus cloud. \citet{rjb04} have found clear evidences for spatial
variations of the aromatic infrared band profiles, likely due to the
spatial variation of the nature of their carriers. A sophisticated analysis
of ISOCAM-CVF data allow them to separate the mid--IR spectra of the
ionized and neutral PAHs from the spectra of carbonaceous very small grains
(possibly PAHs aggregates). The very small grains are located at a larger
distance from the illuminating stars than the PAHs, lending support to the
idea that PAHs are produced from the photo--evaporation of these very small
grains.  While more examples are needed for understanding the origin and
fate of interstellar PAHs, it appears nonetheless that these macro
molecules are released in the gas phase in the UV--illuminated regions of
the interstellar medium, i.e. in the diffuse clouds, in PDRs, etc. In those
regions, the destruction of the carbon skeleton is the main process
limiting the smallest possible PAH size. It is likely that some carbon
bearing molecules are released in the gas phase in the UV--illuminated
regions, either as a secondary product of the evaporation of the dust
particles giving rise to PAHs, or as products of the destruction of the PAH
carbon skeleton.

\section{Summary and Conclusions}
\label{sec:conclusions}

We have presented maps of the edge of the Horsehead nebula in rotational
lines of excited \HH{}, CO, \CeiO{} and simple hydrocarbon molecules,
\CCH{}, \cCCCHH{} and \CCCCH{} with $6''$ resolution. All the hydrocarbon
maps are strikingly similar to each other, and to the mid--IR emission
mapped by ISOCAM~\citep{ab03} while we measured a 3 to $5''$ offset between
the hydrocarbon and \HH{} peaks. State-of-the-art chemical models fail to
reproduce both the PDR hydrocarbon stratification and the absolute
abundances of 2 over 3 observed hydrocarbons. We have examined three
hypotheses for improving the models, and we conclude that the most likely
explanation is we are witnessing the fragmentation of PAHs in the intense
far--UV radiation due to $\sigma$Ori.

A detailed modeling of the chemistry including this new mechanism is beyond
the scope of this paper.  Indeed, such a modeling requires rates for both
the growth (by addition of molecules or of carbon and hydrogen atoms) and
the fragmentation of PAHs.  This last item requires an accurate description
of the fragmentation cascade of PAHs, in all their possible equilibrium
states (ionized, neutral, partially or totally dehydrogenated, \ldots{}).
Laboratory experiments such as the ion cyclotronic resonance cell PIRENEA
in Toulouse~\citep{job03} are key instruments in this perspective.  In
addition, the rate files used by the model need to be updated, with a
special mention on the photo--dissociation rates of the simple carbon
chains.  A critical review of the role of neutral--neutral reactions in
interstellar chemistry is also warranted.

\begin{acknowledgements}
  We are grateful to the \IRAM{} staff at Plateau de Bure, Grenoble and
  Pico Veleta for competent help with the observations and data reduction.
  \IRAM{} is supported by the INSU/CNRS (France), MPG (Germany) and IGN
  (Spain).  This work has benefited from many discussions with C.~Joblin
  and C.M.~Walmsley.  We thank D.~Lis for the communication of the [CI] map
  of the Horsehead nebula in advance of publication. We also thank
  E.~Herbst for providing an updated chemical rate file. MG thanks the
  hospitality of the \CSO{} office in Hilo where she worked on this paper.
  We acknowledge funding by the French CNRS/PCMI program. Finally, we thank
  the referee, J.~Black, for his insightful comments which improved the
  presentation and the discussion of our results.
\end{acknowledgements}

\bibliography{ms1170}%
\bibliographystyle{apj}%

\end{document}